
  \magnification=\magstep1
\settabs 18 \columns
\hsize=16truecm

\def\btd{\bigtriangledown}

\def\b{\bigskip}
\def\bb{\bigskip\bigskip}

\def\no{\noindent}
\def\r{\rightline}
\def\ce{\centerline}
\def\ve{\vfill\eject}

\def\r{\rightline}

\def\L{{\cal L}}

\def\harr#1#2{\smash{\mathop{\hbox to .25 in{\rightarrowfill}}
  \limits^{\scriptstyle#1}_{\scriptstyle#2}}}

\def\R{{\cal R}}

\def\today{\ifcase\month\or January\or February\or March\or April\or
May\or June\or July\or
August\or September\or October\or November\or  December\fi
\space\number\day, \number\year }

\r \today
\bb\bb\bb

\def\DD{\vec \bigtriangledown}


\def\Rrm{\hbox{\rm I\hskip -2pt R}}

\def\e{\rm e}
\def\d{\delta}
\def\p{\partial}

\def\sqr#1#2{{\vcenter{\vbox{\hrule height.#2pt
\hbox{\vrule width.#2pt height#2pt \kern#2pt
\vrule width.#2pt}
\hrule height.#2pt}}}}

  \def\1/2{{\scriptstyle{1\over 2}}}
  \def\a/2{{\scriptstyle{3\over 2}}}
  \def\5/2{{\scriptstyle{5\over 2}}}
  \def\7/2{{\scriptstyle{7\over 2}}}
  \def\3/4{{\scriptstyle{3\over 4}}}

\font\steptwo=cmb10 scaled\magstep2
\font\stepthree=cmb10 scaled\magstep4
\magnification=\magstep1

\def\sqr#1#2{{\vcenter{\vbox{\hrule height.#2pt
\hbox{\vrule width.#2pt height#2pt \kern#2pt
\vrule width.#2pt}
\hrule height.#2pt}}}}

\def \r{\rightarrow}
\def\M{{\cal M}}

\b
\def\DD{{\bigtriangledown}} 

   {\ce {\steptwo   Heat and Gravitation. II. Stability
 }}

   \ce{Christian Fr\o nsdal}
\b
   \ce{\it Physics Department, University of California, Los Angeles CA
   90095-1547 USA}
  \b

\def\sqr#1#2{{\vcenter{\vbox{\hrule height.#2pt
\hbox{\vrule width.#2pt height#2pt \kern#2pt
\vrule width.#2pt}
\hrule height.#2pt}}}}

\def \r{\rightarrow}
\def\M{{\cal M}}

\b
   \no ABSTRACT ~  Some features of hydro- and thermodynamics, as applied to
atmospheres and to stellar structures, are puzzling:  1.  The suggestion,
first made by Laplace, that our atmosphere has an adiabatic temperature
distribution, is confirmed for the lower layers, but the reason why 
it should be
so is difficult to understand.     2. The
standard treatment of relativistic thermodynamics does not allow for a
systematic treatment of mixtures, such as the mixture of  a perfect gas with
radiation. 3. The concept of mass in applications of  general relativity to
stellar structures is less than completely satisfactory.  4.
Arguments in which a concept of energy plays a role, in the context of
hydro-thermodynamical systems and gravitation, are not always convincing.
   It was proposed that a formulation of
thermodynamics as an action principle may be a suitable approach to adopt for a
new investigation of these matters.

In this second article of a series we propose to base criteria of stability 
on the 
hamiltonian functional that is provided by the variational principle, 
to replace the reliance that has often been placed on {\it ad hoc}
definitions of the ``energy".  We introduce a new virial principle that is formulated
entirely within  the Eulerian description of hydrodynamics, 
which allows a simpler derivation of a well known stability criterion
for  polytropic stellar configurations. Boundary conditions are based
entirely on mass conservation.

The new approach  is tested on isothermal and
polytropic atmospheres and then used to initiate a new study of stars.
Traditional results for polytropic, spherical configurations are confirmed,
but our study gives new insight and new results for the case that radiation
is taken into account. 
\b

PACS Keywords: Atmosphere, astrophysics, action principle, virial theorem.
   \ve
\no{\steptwo I. Introduction}

Stability of a dynamical system that consists of a gas held together 
by mutual 
gravitational  forces can be studied by comparing the energy of a presumptive 
bound configuration with
that of an infinitely diffuse state.  But a necessary condition, for such 
comparison to carry any weight, is that a framework exists within which a
well defined hamiltonian functional of the dynamical variables can be provided.

The literature contains    stability studies where
this condition,  if not clearly stated,  was nevertheless satisfied. 
But there are numerous instances, even in textbooks,  where  an expression
for the energy is written down without any supportive dynamical framework.

In these papers we insist on basing all investigations on a variational 
principle.
The dynamical equations are in most cases the same as those of previous
studies,  but in the case that the effect of radiation is included  we 
are breaking new ground.

The paper begins with a very brief summary of the first paper of this 
series (Fronsdal 08).
A well known variational principle for irrotational hydrodynamics is
expanded by elevating the temperature to the role of an independent
dynamical variable. After a brief summary of standard atmospheric models we
introduce a new ``virial"  theorem. It is an analogue of the classical
virial  theorem,  in which the particle coordinate operator $\vec r \cdot
\vec p$ is replaced by
$\rho\Phi$, where $\rho$ is a mass density and $\Phi$ is the velocity
potential  (the canonical conjugate of $\rho$).
Using this new virial theorem in a manner that recalls Jacobi's's original 
use of the classical analogue
we are led to the well known stability criterion ($n <3$) for 
polytropic atmospheres.
No appeal is made to kinetic theory and no approximations are made.
 
Section 2 investigates the stability of several atmospheric models. The
boundary conditions, that are central to the question of stability, are
deduced in each case from the conservation of mass. This is a reminder of
the fact that the usual approach to relativistic thermodynamics abandons the
continuity equation of classical hydrodynamics, a defect that can easily be
repaired (Fronsdal 2007) .

We study the stability of several atmospheric models by  perturbation theory.  The important 
role of mass conservation in the choice of initial conditions is emphasized. 

This paper is limited to the non relativistic approximation. The relativistic case will be examined in another paper of this series.

\b\b
\ce{\bf 1.1. Hydrodynamics}

Basic hydrodynamics deals with a density field
$\rho$ and a velocity field $\vec v$ over $\Rrm^3$, subject to two
fundamental equations, the equation of continuity,
$$
\dot \rho + {\rm div}(\rho \vec v) = 0,~~ \dot \rho := {\p\rho\over \p
t},\eqno(1.1)
$$
and the hydrodynamical equation (Bernoulli 1738)
$$
-{\rm grad}~ p = \rho{D\over Dt}\vec v := \rho(\dot{\vec v} + \vec v\cdot
{\rm grad} ~\vec v).\eqno(1.2)
$$
This involves another  field, the scalar field $p$, interpreted as the
local pressure.

We  assume that the velocity field can be
represented as the gradient of a scalar field,
$$
\vec v = -{\rm grad}~ \Phi.\eqno(1.3)
$$
In this case the hydrodynamical condition is reduced to
$$
    {\rm grad} ~p = \rho~{\rm grad}~ (\dot\Phi - \vec
v^2/2).\eqno(1.4)
$$
To complete this system one needs a relation between the fields $p$ and
$\rho$.

A local functional $V[\rho]$ is defined up to an additive  linear term by 
$$
p = \rho V' - V,~~ V' := dV/d\rho.\eqno(1.5)
$$
Then $dp = \rho~ dV'$ and the equation (1.4) becomes, if $\rho
\neq 0$,
$$
V' =  \dot \Phi -  \vec v^2/2 + \lambda,~~ \lambda ~
{\rm constant}.\eqno(1.6)
$$
  
The two  fundamental equations that characterize irrotational  hydrodynamics 
are
$$\eqalign{&
\dot \rho + {\rm div}(\rho \vec v) = 0,~~ \dot \rho := \p\rho/\p t, 
 \cr &
  \dot \Phi -  \vec v^2/2 + \lambda= \p V/\p\rho ,\cr}  \eqno(1.7)
$$
together with the defining equations
$
\vec v = -{\rm grad}~ \Phi,~~ p := \rho V' - V. 
$
These equations are the Euler-Lagrange
equations associated with the action (Fetter and Walecka  1980)
$$
A[\rho,\Phi] = \int dtd^3x ~{\cal L},~~ {\cal L} =  \rho(\dot\Phi - 
\vec v^2/2 +
\lambda) - V[\rho].\eqno(1.8)
$$
{\it The value of this last circumstance in the present context lies in the fact that it  gives us a
valid concept of a total energy functional.}

\b\b

 \ce{\bf  1.2. The mass}
  The conserved density $\rho$
will be taken to have the interpretation of mass density, and the total
mass is the constant of the motion
$$
M = \int d^3 x ~\rho.
$$
 
Since the total mass is a constant of the motion it is natural to fix
it in advance and to vary the action subject to the constraint
$\int  d^3x\, \rho(x) = M$. The parameter $\lambda$ takes on 
the role of  a
Lagrange multiplier and the action takes the form
$$
A = \int  d^3x\Big(\rho(\dot\Phi - \vec v^2/2-\phi) -V\Big)
+\lambda\Big(\int   d^3x \rho - M\Big).\eqno(1.9)
$$
The  gravitational potential $\phi$ is included.

The conservation of mass has important implications for boundary
conditions.
\ve

\ce{\bf 1.3. Isothermal and polytropic atmospheres}

An ideal gas at equilibrium, with constant temperature, obeys the gas law
$$
p/\rho = \R T.\eqno(1.10)
$$
Pressure and density are in cgs units and
$
\mu \R = .8314\times 10^8~erg/K ,
$
  where $\mu$ is the
atomic weight.     

 \b
In hydrodynamics,    the isothermal atmosphere can be given a lagrangian
treatment by taking
$$
V = \R T \rho \log\rho.\eqno(1.11)
$$
  We
suppose that the gas is confined to  the section  $z_0<z<z_0 + h$ of a
vertical cylinder with base area ${\cal A}$ and expect the density to fall
off at higher altitudes.  The action density, for a
 gas at constant temperature $T$ in a constant gravitational field $\phi
= gz$, $g$ constant, is
$$
{\cal L}[\Phi,\rho] = \rho\,(\dot\Phi - \vec v^2/2  - gz + \lambda) -
{\cal R}T\rho\log\rho.\eqno(1.12)
$$
We may consider this an isolated system with fixed mass.  

At equilibrium $\dot\Phi = 0, \vec v = 0, \dot \rho = 0$ and the
equation of motion is
$ V' = {\cal R}T(1 + \log\rho) = \lambda - gz,$
hence
$$
\rho(x,y,z) = \e^{ -1+\lambda/\R T}\e^{-gz/\R T},~~ M = {\cal A}{ {\R}
T\over g}\e^{-1 +\lambda/\R T}(1- \e^{-gh/RT})~\e^{-gz_0/\R T}
$$
and after elimination of $\lambda$
$$
\rho = {gM\over {\cal A}{\R} T}{\e^{-g(z-z_0)/\R T}\over 1-
\e^{-gh/RT} },~~ p={gM\over {\cal A}}~{\e^{-g(z-z_0)/\R T}\over 1-
\e^{-gh/RT}}.\eqno(1.13)
$$
(If the atmosphere extends to infinity the pressure at the bottom is $gM/{\cal A}$, as it shoul be.)
A difficulty with this model is that the specific internal energy
is $\R T\log\rho$, which is very different from that of an ideal gas.

The isothermal atmosphere
is usually abandoned in favor of the polytropic atmosphere.  
A polytropic gas can be described by the lagrangian (1.9), with
$$
V = \hat  a\rho^{\gamma}, ~~ \hat a, \gamma ~{\rm constant}.
$$
Variation with respect to $\rho$  gives
$$
p ={\hat a\over n}\rho^{\gamma},~~{1\over n} = \gamma-1.
$$
The temperature does not appear explicitly but is taken to be determined by the
gas law,  Eq.(1.10).  
At mechanical equilibrium $\vec v = 0, \dot \rho =
0$ and $\lambda - gz= \hat a\gamma\rho^{1/n}$, hence
$$
   \rho =
({\lambda - gz\over \hat a\gamma})^n.
$$
Since the density must be
positive one does not fix the volume but assumes that the atmosphere
ends at the point $z_1 = \lambda/g$. Then
$$
  M= {{\cal A}}({g\over
\hat a\gamma })^n\int_{z_0}^{z_1}(z_1-z)^ndz = {{\cal A}h\over
n+1}({gh\over \hat a\gamma})^n. \eqno
$$
This fixes  $h$ and thus $z_1$
and $ \lambda$. If the atmosphere is an ideal gas then
  the temperature varies with altitude according
to
$$
{\cal R}T = p/\rho = {\hat a\over n}\rho^{1/n} = g{ z_1-z\over n+1}
\eqno(1.14)
$$
 
For air, with
atomic weight 29, $\R = 2.87\times 10^6 erg/gK$ and $n = 2.5.$  At
sea level,\break$g = 980 cm/sec^2$,  the density  is   $\rho  =1.2
\times 10^{-3}  g/cm^3$, the pressure $ p =1.013 \times 10^6
dyn/cm^2$.  Thus
$
p/\rho = .844 \times 10^9 cm^2/sec^2, ~~ T = T_0
= 294K,~~ z_1 = 3.014\times 10^6 cm \approx 30km.
$ 
and the dry lapse
rate at low altitudes is  $ -T'=294/ z_1= 9.75  K/km.$  

The specific internal energy of this model is $np/\rho = n\R T$, as it
should be for an ideal gas. However, the temperature is not treated
as an independent dynamical variable, instead it is fixed by the constraint
(1.10). The theory developed so far is partially ``on shell".

A variational ,principle is incomplete without the specification of boundary 
conditions. The only boundary condition that we shall apply systematically
is the conservation of mass, which controls the material flow  
at the boundary.

\b\b
 \no{\steptwo II. Thermodynamics}
  
\ce{\bf 2.1. The adiabatic lagrangian}
   The  action principle can be extended so as to treat
  the temperature as an independent dynamical variable.
  
  Two kinds of additions can be made to the lagrangian (1.9)
    without spoiling the equations of motion that
    are essential to hydrodynamics. To preserve the equations of motion as 
well as the correct expression for the internal energy, we  have been led to
the lagrangian for an ideal gas,
$$
{\cal L}[\Phi,\rho,T] = \rho(\dot\Phi - \vec v^2/2 -\phi + \lambda )
    -{\cal R}T\rho\log {k\over k_0}  +{a\over 3}T^4,\eqno(2.1)
$$
with $k := \rho/T^n$. The last term is the contribution of radiation, 
assumed as is usual to be that of a black body; the constant $a$ is the
Stefan-Boltzmann constant, $a = 7.56\times 10^{-15} ergs/cm^3K^4$.

Variation of the velocity potential $\Phi$ gives the continuity equation 
as before.
Variation of the density $\rho$ leads to
$$
\dot\Phi - \vec v\,^2/2-\phi + \lambda = \R T (1+\log {k\over
k_0}).\eqno(2.2)
$$
Taking the gradient one gets
$$
\rho{D\over Dt}\vec v -\rho {\rm grad }~\phi= - {\rm grad~}   p,\eqno(2.3)
$$
where $D/Dt = \p/\p t +\DD\cdot \vec v$ is the total derivative time and 
$$
 p = \R T\rho + {a\over 3}T^4.\eqno(2.4)
$$
 Variation with respect to $T$ gives
$$
\R \Big(n- \log {k\over k_0}\Big) \rho  
+{4a\over 3} T^3 = 0, \eqno(2.5)
$$  
With the neglect of the radiation term, or if $n = 3$,  
this reduces to the polytropic equation of change, $k = $ constant.

For conditions as in the atmosphere of the Earth,  
$\log (k/k_0) - n \approx 10^{-10}$. When the emphasis is not on
applications we shall use units such that $k_0 = 1$.

The on shell hamiltonian  density, obtained from (2.1) with the help of
(2.5),  takes the form
$$
h = \rho \vec v^2/2 + \rho\phi + u,~~ u = 
n\R T\rho + aT^4.\eqno(2.6)
$$
 
  We suggest that, when $n \neq 3$,  
using the lagrangian (2.1) is preferable to the usual assumption 
$\beta := p_{\rm gas}/p_{\rm tot}$ = constant, which is true only when $n
= 3$.

  {\it We have  an action principle, with dynamical variables
  $\rho$
and $T$, that   reproduces all of the equations that characterize the
equilibrium configurations,   including the contribution of the photon gas to the energy and the pressure, as well as the
standard, hydrodynamical relations of an ideal gas.}

\b\b

  \ce{\bf 2.2. Virial theorem}

Both (2.4) and (2.6) are usually derived from considerations outside
the proper domain of thermodynamics. We  prefer
an axiomatic foundation of thermodynamics that is complete in the
sense that it does not need other input. As an  example
let us discuss the use of the virial theorem to make certain
predictions concerning stability.

The virial theorem was introduced into the present context by Kelvin. It
is based on the scaling properties of the hamiltonian of a system of
particles. If
$H = K+V$, kinetic energy plus potential energy, then the lagrangian is
$K-V$ and the equations of motion imply that, up to a time derivative,
$$
 {d\over dt}(\vec r\cdot \vec p) = 2K - \vec r\cdot \DD V,
$$
where $\vec p = m\vec v, K$ is the kinetic energy $m\vec v\,^2/2$ and $V$ 
is the potential. If the potential is homogeneous of degree $n$ then
$$
{d\over dt}(\vec r\cdot   p) = 2K-nV,
$$
If the particle goes through a cycle then the average of this quantity 
over the cycle is zero. 
In the case examined by Kelvin the potential is homogeneous of degree -1,
so that, \underbar{in the case of periodic motion}, when average is taken
over a period, $V = -2K$.  This is used
to prove stability, since the total energy is $E = K + V = -K$ is
negative. Actually, what is proved is that, if the motion is periodic
then it is bounded. 

The application to continuous systems, developed by Clausius (1851),  is much
more difficul. According to  Chandrasekhar
  (1938) (pp. 49-51), who also quotes Poincar\'e, the internal energy
is the kinetic energy associated with  the
microscopic motion of the molecules. It is assumed, usually without
discussion, that the presence of gravitational forces do not affect the
internal energy, and that the total energy is obtained by simply adding the
gravitational potential energy to it. In the present approach we are led 
to the same conclusion, 
the energy is identified with the hamiltonian. But the theory does not 
authorize drawing on kinetic theory to formulate a virial theorem. In fact,
some attempts to do so have been successful only after making various
assumption and/or approximations. See for example Collins  (2007).

There is; however, a virial theorem associated with a lagrangian of
the type (2.1), that we abbreviate as
$$
{\cal L} = \rho(\dot \Phi - \vec v^2/2 ) - \hat V.
$$
(The potential $\hat V$ includes the gravitational field and the lagrange multiplier.) Variation of
$\Phi$ and of $\rho$ gives the equations of motion
$$
\dot \Phi = \vec v^2/2 +(d\hat V/d\rho),~~ \dot\rho = -{\rm div}(\rho
\vec v),
$$
In accord with the boundary condition discussed at the end of Section 1.3 (the conservation of mass), 
we drop the boundary term $-\int\Phi\rho\vec v\cdot \vec d\sigma$ to get
$$
\int d^3x {d\over dt}(\rho\Phi) = \int d^3x\Big(\rho{d\hat V\over d\rho} -
\rho\vec v\,^2/2\Big).
$$
If the system goes through a cycle then the average of this
quantity over the cycle is zero,
$$
 \int dx~\rho\vec v\,^2/2~ = \int dx~ \rho{d\hat V\over
d\rho}.\eqno(2.7)
$$
  In the case of the lagrangian (2.1), but neglecting the radiation term,  we obtain, 
 $$
 \int dx\Big(\rho\vec v\,^2/2  +\rho(\phi-\lambda) +
  \R T\rho(1+\log k) \Big) = 0.\eqno(2.8)
$$

To draw conclusions from this result we adopt the strategy of Jacobi (1889) ,
noting that the total energy  is
$$
E = \int d^3x\Big(\rho(\vec v\,^2/2 + \phi/2) + \R T\rho\log k\Big).
$$
By replacing the gravitational potential $\phi$ by $\phi/2$ we are 
specializing to the case that the interaction is between the particles, not
with a fixed source. In this case the interaction energy is bilinear in the
density and variation of $\rho$ introduces a factor of 2. The equations of
motion are always the same but the factor 1/2 is needed in the action.

Combining the last two equations we obtain
$$
\gamma E = \int d^3 x\Big({1\over n}\rho\vec v\,^2/2 + 
({1\over n}-1)\rho\phi/2 + \lambda\rho\Big).
$$
The last two terms can be evaluated. 
The calculation was done first by Ritter (1870); it is in many textbooks,
and it is repeated later in this paper.  The first result,
$$
\int d^3 x \lambda \rho = \lambda M = \lambda  \phi(R) = -{GM^2\over R},
$$
is obtained with the help of the boundary condition at the surface and
elementary properties of the gravitational interaction. The derivation of 
the following formula,
$$
\int d^3x \rho\phi/2 = E_g = {-3\over 5-n}{M^2G\over R}.
$$
makes use the known degree of homogeneity ($\gamma$) of the heat term with
repect to the density.  It is important that the equation of motion was
not used. The sum is
$$
\int d^3 x\Big(({1\over n}-1)\rho\phi/2 + \lambda\rho
\Big) = {(n-3)(n+1)\over (5-n)n} {GM^2\over R}. 
$$
This shows that, when $n > 3$, the potential energy near the point of 
equilibrium is positive, which allows the gas to diffuse;  the configuration
is unstable. If $n=3$ there is a stable solution with zero
energy, and another solution for which the energy is not defined. It just
walks away, like a free particle or a critically damped harmonic oscillator.

\b\b\b\b

\no{\steptwo  III. Stability of some atmospheres}

In all the examples of fluctuations around a static solution that follow
we consider only adiabatic fluctuations; in the isothermal case the
temperature is fixed and in the polytropic case the fluctuations of $T$
are governed by $\delta \rho/\rho = n\delta T/T$,   by fixing $k =
\rho/T^n$. One has to ask if this limitation to adiabatic fluctuations is
fully justified. To study this question it is necessary to introduce
entropy and to consider the heat equation, as we have attempted to do in
the first paper (Fronsdal 2009). As pointed out by Emden (1907), it is
necessary to assume that the heat flow has zero divergence, and this is
supported by the fact that the polytropic atmosphere has constant lapse
rate.

. 
\b
\ce{\bf 3.1. The isothermal  column}

The equations of motion (1.7), with $V = \R T\rho\log\rho$ (see Eq.(1.12))
are
$$
\dot\rho + (\rho v)' = 0,~~ \dot v +\big(v^2/2 +\R T (1+\log\rho)\big)' =
0.
$$
The prime indicates derivation with respect to $z$.
We consider the space that is tangent to a static solution with density
$\rho_0$.  Setting
  $
\rho = \rho_0 + \delta\rho
  $
we have the following equations for the perturbation $\delta\rho$,
$$
 \delta\dot\rho +(\rho_0 v)',~~\dot v +  \R T (\delta\rho/\rho_0)'=0,
$$
  Thus
$$
\delta\ddot\rho = \R T(\rho_0\alpha')',~~ \alpha :=
\delta\rho/\rho_0. 
$$
For a harmonic mode with frequency $\omega$,
$$
-\omega^2\delta\rho =  R T (\rho_0\alpha')',\eqno(3.1)
$$
 and
$$
-{\omega^2\over \R T}\int \alpha^2\rho_0dz = \int\alpha(\rho_0\alpha')'dz
  =  \rho_0\,\alpha\alpha'\Big|_0^\infty - \int \rho_o \alpha'^2 dz.
$$
The configuration is stable if this implies that $\omega^2>0$, which will
  be the case if the boundary term vanishes. To justify any choice of
boundary conditions we have only the conservation of mass,
$\int\delta \rho dz = 0$. This ensures that $\delta\rho$ fall off at
  infinity and we are left with $-\delta\rho(0)
\alpha'(0)$.

We shall show that $\alpha'(0) = 0$. Eq.(3.1) tells us that
$$
-{\omega^2\over \R T}\alpha =    (\rho'_0/\rho_0)\alpha' + \alpha''
= -{g\over \R T} \alpha' + \alpha''.
$$
This is a linear differential equation with constant coefficients,
with general solution
$$
\alpha(z) =  A\e^{k_+z} + B\e^{k_-z},~~ k_\pm =
{g\over 2\R T} \pm \sqrt{({g\over 2\R T})^2 - {\omega^2\over \R T}}.
$$
Since, up to an irrelevant constant factor, $\rho_0 = \exp(-gz/\R T)$,
$$
\delta\rho = \rho_0\alpha =  A\e^{a_+z} + B\e^{a_-z},~~ a_\pm =  - {g\over
2\R T}\pm \sqrt{({g\over 2\R T})^2 - {\omega^2\over \R T}}.
$$
In order that $\delta M = {\cal A}\int \delta\rho dz$ vanish we need for the
coefficients $A,B$ to be both non zero, and convergence of the integral then
requires that $\omega^2 > 0$. In this case 
$$
\delta M = \int\delta\rho dz = -{A\over a_+} - {B\over a_-} = {A\over k_-} +
{B\over k_+},
$$
the vanishing of which requires that $\alpha'(0) = 0$.  Therefore, not
only is the condition $\omega^2 >0$ verified; it is  also confirmed that the
boundary condition $\alpha'(0) = 0$ is the only one possible. We have seen
that this choice of boundary conditions is the one that ensures the
conservation of mass.  

The fluctuation $
\delta\rho$ does not vanish at the lower end, but the velocity does, as
is natural. In fact, we can reach the same conclusion more easily as
follows. Conservation of the current implies that
$$
\rho v\Big|_0^\infty = 0.
$$ 
Since, as we showed, the upper limit cannot give a contribution,
the velocity must vanish at the lower limit.

In this model, and in the one examined next, stability it physically obvious.
The virial theorem has no interesting application in either case.

\b\b

\ce{\bf 3.2. The polytropic column}

Let us leave the parameter $k =\rho/T^n$ free and fix the value of $n$. This
conforms to the usual approach when the temperature is fixed by edict, but
it is consistent with our formulation if $n = 3$ only, or in the case that
the effect of radiation is neglected. We study the stability to vertical
perturbations.

The static solution of (2.2) is
$$
cT = \lambda-gz,~~ c := \R(1+\log k).
$$
A first order perturbation satisfies
$$
\dot\Phi + \delta\lambda = c\delta T,~~{\rm thus}~~\dot v =- c\,\delta
T'.\eqno(3.2)
$$
The equation of continuity gives
$$
\dot\rho = -({v\,\rho})',~~ \ddot \rho = -(\dot v \rho)'.
$$
Let $x = \lambda/g - z,~ 0<x<\lambda/g$ and let $f' = df/dx$ from now on.
Solutions of the type $\delta T = \exp(i\omega t)f(x)$
satisfy the equation
$$
(x^n\delta T')' + {\nu^2\over x}(x^n\delta T) = 0, ~~ \nu^2 =
n\omega^2/g.\eqno(3.3)
$$
The solution that is regular at the origin of $x$ (the top of the
atmosphere) is
$$
\delta T = \,  _0\hskip-.5mmF_1(n,-\nu^2x)\e^{i\omega t}.
$$
(Another solution may be obtained by the transformation $n \rightarrow 2-n,
x\rightarrow 1/x$.) 
    The generalized hypergeometric function is
positive for positive argument and it oscillates around zero for negative
argument.

From (3.3) one obtains
$$
\nu^2\int x^{n-1}(\delta T)^2 dz =- \int \delta T (x^n\delta T')' =
-x^n\delta T\delta T'\Big|_0^{\lambda/g} + \int x^n(\delta T')^2.
$$

\b
\no\underbar{Boundary conditions}. If we fix $\delta T = 0$ at the bottom of
the column the boundary term vanishes, 
provided that the upper limit makes no contribution, and in this case the mass would be preserved. There are solutions for $\nu^2>0$ only, oscillatory in time.  

   The natural boundary condition is the preservation of the mass,
thus
$$
\delta M = \int\delta\rho dx = nk\int T^{n-1}\delta Tdx = 0.
$$
This may happen for a discrete set of positive values of $\nu^2$.
For negative values of $\nu^2$ the integrand is definite so that it can not
happen. The calculation is valid only in the case $n = 3$ (for all $n$  when
the effect of radiation is neglected); this atmosphere is stable.

   The problem can be
converted to a standard boundary value problem by rescaling of the
coordinate.

\b

\ce{\bf III. The polytropic gas sphere. The hamiltonian}

Here we study the self gravitating polytropic gas. A correction is needed
  in the expression for the lagrangian, and we need to take care with respect
to the definition of the gravitational potential.

The gravitational energy of a system in mutual interaction is the
following functional of the mass density  $\rho$,
  $$
E_g[\rho] = -{G\over 2} \int d^3xd^3x'.
{\rho(\vec x)\rho(\vec x\,')\over |\vec x - \vec x'|}.\eqno(3.4)
$$
It vanishes in the limit that the gas is diffused over an infinite volume
and it is negative for all other density profiles.

This potential energy contributes to the hamiltonian so that we have to
include a term $-E_g[\rho]$ in the lagrangian. It contributes to the
equation of motion that comes from variation of $\rho$ a term
$-\rho\phi$, as in the other cases examined, but now
$$
\phi[\rho](\vec x) = \phi(\vec x) = -\int d^3x
{G\rho(\vec x\,')\over |\vec x- \vec x\,'|}.\eqno(3.5)
$$

Our text books show how to evaluate the potential in the case of a
spherically symmetric distribution, when one writes $\rho(x)\rightarrow
\rho(r),~ \phi(x)\rightarrow \phi(r)$, with
$$
\phi(r) = -G\int{\rho(r')\over z} r'^2\sin\theta'dr'd\theta'd\phi',~~ z =
r^2 + r'^2 - 2rr'\cos\theta',~~ z> 0.
$$
The trick is to replace $(r',\theta')$ by (r',z) as independent variables
of intgration, using
$$
zdz = rr'\sin\theta' d\theta'.
$$
The integral becomes
$$
\phi(r) = -2\pi G\int\rho(r'){r'\over r}dr'dz = -2\pi G(I_+ + I_-),
$$
where $I_+$ is the contribution from the region $r'<r$ and $I_-$ is the rest. Now
$$
I_+ = {1\over r}\int_{r-r'}^{r+r'}\rho(r')dr'dz = {2\over r}\int_0^r\rho(r')r'^2dr',
$$
$$
I_- = {1\over r}\int_{r'-r}^{r+r'}\rho(r')dr'dz =
2\int_r^\infty\rho(r')r'dr',
$$
It results from this that
$$
\phi'(r) = 4\pi G{1\over r^2}\int_0^r\rho(r')r'^2dr'.
$$
the mass contained within the distance $r$ from the center is
$$
{\cal M}(r) = \int_{r'<r}\rho(x')d^3x',
$$ and the result of the calculation is that
$$
\phi'(r)  = {G{\cal M}\over r^2},~~ \phi(R) = -{G\M\over R}.
$$
The potential is negative, increasing from the origin to the boundary 
and on towards zero at infinity,
since for $r>R, \phi(r) = -G\M/r$.

We shall need 3 important integral forrmulas, the first two  are found by
by integration by parts,
$$
E_g = -4\pi\int_0^R{1\over 2}\rho\phi r^2dr = {1\over 2}\int\M'\phi dr =
{1\over 2}M\phi(R) - {1\over 2}\int\M\phi'dr,\eqno(3.6)
$$
$$
-{1\over 2}\int \M \phi' dr = -{1\over 2}\int{G \M^2\over r^2}dr =
{1\over 2}\int G \M^2({1\over r})'dr = {1\over 2}{GM^2\over R} - \int{G\over
r}\M d\M.\eqno(3.7)
$$

 The third is obtained by Eddington (1926) 
by  using the polytropic relation between $T$ and $\rho$.
 We need the equation of motion,
 $$
 \lambda -\phi = cT,~~ c = \R(\log{k\over k_0} + {n\over 3}),
 $$
but only to extract the boundary conditions.
 Since $T(R) = 0$ it tells us that $\lambda = \phi(R)$. This field, namely
 $$
 \psi := \phi(R) - \phi,
 $$ 
 is Eddington's gravitational potential. 
The polytropic relation is seen as relating $\psi$ and $\rho$ and gives the
result, here expressed in terms of $\phi$, is
$$
\int{G\over r}\M d\M = {3\over n+1}\int\M\phi'dr'.\eqno(3.8)
$$
Combining (2) and (3) we obtain the data that will allow us to determine 
the value of the hamiltonian,
$$
\int\M \phi' dr  =  {n+1\over 5-n}{GM^3\over R},~~ \int {G\over r}\M d\M
= {3\over 5-n}{GM^2\over R} 
$$
and
$$
E_g = -{3\over 5-n}{GM^2
\over R}.\eqno(3.9)
$$

The hamiltonian density is
$$
h = \rho\vec v\,^2/2 + {1\over 2}\rho\phi + \R T\rho\log k + {a\over 3}T^4,
$$
or, in view of (2.5),
$$
h ={1\over 2}\vec v\,^2 + {1\over 2}\rho\phi + {3c\over 4}T\rho,~~ c =
\R({n\over 3} + \log k),
$$
whence the hamiltonian (= total energy) is
$$
H = \int d^3x h = E_g + \int d^3x({1\over 2}\vec v\,^2 + {3c\over
4}T\rho).\eqno(3.10)
$$
Finally, the equation of motion $\lambda-\phi = cT$ yields
$$
\int cTd^3x = \lambda M +2E_g = {n+1\over 5-n}{GM^2\over R} 
$$
 Hence (4.12) reduces, in the static case, to
$$
H = -{3\over 5-n}{GM^2
\over R} + {3\over 4}{n+1\over 5-n}{GM^2\over R} =  {3\over 4}{n-3\over
5-n}{GM^2\over R}.
\eqno(3.11)
$$
The conclusion that is drawn from this formula is that the equilibrium
configuration is stable only if $n < 3$. 

\b

\no{\bf Remark.} It is useful to calculate the relative contribution of
radiation to the total energy; this can be done exactly when $n = 3$,
using the equation of motion (2.5),
$$
\R k(\log {k\over k_0} -3) = {4a\over 3}.
$$ 
We already gave a numerical example: in the case of the Earthly atmosphere
  $\log(k/k_o)-n \approx 10^{-10}$. In general, the internal energy,
with radiation included, can be expressed on shell as
$$
u = n\R T\rho + {a\over 3}T^4 =: u_{gas} + u_{rad}.
$$
and it is natural to ascribe the first term to the gas and the other to
radiation. Defining the quantity $\beta$ as Eddington does, namely
$$
\beta = p_{gas}/(p_{gas} + p_{rad}) = u_{gas}/(u_{gas} + u_{rad}),
$$ we obtain
$$
\beta = {\log(k/k_0)-3\over \log(k/k_0) + 1},
$$
which in the numerical example is approximately $.25\times 10^{-10} $.
\b

\ce{\bf Historical review. }
 
According to Kelvin (1800), the conclusion that $n = 3$ is critical for
stability was announced by Perry, but Perrys paper could
not be found. The same result was obtained, in the same year, by Ritter
(1880) and Betty (1880). Ritter's calculation has often been repeated,
but the original is by far the most coherent; it is very close to
the one that we have given above. 

Ritter does not formulate a lagrangian or a hamiltonian dynamics. He
defines the total energy as the sum of the internal energy and the
gravitational energy and postulates that this quantity is conserved.  
He did not have a dynamical theory in which the total
energy is a well defined functional of the dynamical variables, but it is
not a surpris e to find that such a dynamical theory exists.

Later writers are much less careful. Eddington repeats all of Ritter's
calculations, including the derivation of (3.8) that we have attributed
to Eddington because his calculation is more easily accessible. But
Eddington does not make use of the expression for the thermodynamical
internal energy; instead he appeals to the interpretation of the gas as
a collection of particles with mutual gravitational interactions. 
His claim that the result applies when radiation is taken into account
is valid only in the case that $n = 3$, as the reader can easily verify.
Some modern writers give Ritter's calculation the attention that it merits
(Weinberg 1972), but some others take shortcuts that make their conclusions
far less compelling.  Kippenenhahn and Weigert (1990) reach the conclusion 
that $n = 3$ is critical on the basis on an expression that is simply
\underbar{defined} to be the total energy.
Chandrasekhar (1958) also discusses this problem.

The above calculation made use of the polytropic relation, so it  is
valid in two cases: for all $n$ if the Stefan-Boltzmann term is
neglected, and in the case $n = 3$ whether the Stefan-Boltzmann term is
included or not. Eddington claims the result for all $n$, but if $n\neq
3$, under the condition that the radiation pressure is a fraction of the
total pressure, fixed throughout the star, which is not verified in our
model, and under the additional condition that the polytropic relation
between temperature and density is valid, which is also not verified.

 All this leaves unsettled the question of stability for the case that the
radiation term is included. However, if there is a unique, critical value
of $n$ then it must be $n = 3$, since Ritter's calculation applies to
this special case, even when the energy of radiation is taken into
account.

\b\b

\ce{\bf IV. The polytropic gas sphere. Stability }
Here we shall invetigate the stability of the polytropic sphere directly, by perturbation theory.

We use the lagrangian
$$
\L = \rho(\dot\Phi -\vec v\,^2/2 - \phi/2 + \lambda) - \R T\rho\log k +
{a\over 3}T^4,~~ k := \rho/T^n.\eqno(4.1)
$$
Variation with respect to $T$ gives
$$
\R(\log k - n) = {4a\over 3}T^3/\rho.\eqno(4.2)
$$
With  $n = 3$ this makes $k$ a constant, and $
\log k = n$ when radiation is neglected. In the remainder of
this section, we set, for all values of $n$,
$$
\rho = k T^n,~~ k ~{\rm constant}.
$$
This is the usual polytropic relation used by Eddington and others, but it
is consistent with (4.1) only when $n = 3$.
  The remaining dynamical equations are
$$
-{D v\over Dt} = \phi' + cT', ~~ \dot\rho + r^{-2}( r^2\rho v)' = 0,
$$
$$
  4\pi G  \rho = r^{-2}(r^2\phi')',~~ \rho = kT^n.
$$
  \b
$\bullet $ The static solution. Eliminate $\phi$ by $\phi' = -cT'$ and change
variables, setting $r = x/\alpha$, $\alpha$ constant, Poisson's equation
becomes
$$
{4\pi G   k\over c\alpha^2} x^2T^n + (x^2T')' = 0,
$$
where the prime now stands for differentiation with repect to $x$. Set
$f(x) = T(x)/T(0)$ and $ \alpha =  \sqrt{4\pi G / cT(0) }$ so that
finallly
$$
x^2 f^n +(x^2 f')' = 0,~~ f(0) = 1,~~ f'(0) = 0.
$$
The solution decreases monotoneously to zero at x = X, this point taken
to be the surface of the star. At the outer limit
$
f(x) \propto X/x-1 +o(X-x)^n.
$
The integration is done easily and accurately by Mathematica, especially so
for integer values of $n$.  The radii are,
for $ n = 2: X = 4.355, ~n = 3: X = 6.89685636197,~n = 4: X = 14.9715.$

$\bullet$ For the fluctuations we assume harmonic time dependence, then the
equations are
$$
-\omega^2 r^2\delta\rho =  \big(r^2\rho\,(\delta\phi' + c\delta T')\big)', ~~
\delta\rho = nkT^{n-1}\delta T,\eqno(4.3)
$$
$$
4\pi G r^2\delta\rho = (r^2\delta\phi')'.\eqno(4.4)
$$
Introduce the function
$\delta
\M = r^2\delta\phi'$. Eq.s(4.3-4) then take the form
$$
-{\omega^2\over 4\pi G} \delta \M  =   \rho\delta \M
+ r^2\rho  c\delta T'  + {\rm constant},
$$
where the constant can only be zero, and
$$
(4\pi G) r^2(nkT^{n-1}\delta T) =\delta\M',
$$
Elimination of $\delta T$ leads to
$$
- {\omega^2\over 4\pi G} \delta \M  =   \rho\delta \M
+ {c\over 4\pi G kn}r^2\rho \Big({\delta \M'\over x^2
T^{n-1}}\Big)'.
$$
Changing the scale as before we get
$$
-\nu^2\delta\M = f^n\delta \M + {1\over n} x^2 f^n\Big({\delta\M'\over x^2
f^{n-1}}\Big)', ~~ \nu^2 = {\omega^2\over 4\pi G k T^n(0)}.\eqno(4.5)
$$

The crucial point is the choice of the correct boundary conditions, at $x =
0$ as well as the outer surface ($x = X$). At the center the solutions
take one of two forms, $1 +  C x^2 + ...$, which is unphysical,  or else
$x^3 + C x^5 + ..$. Accordingly we set
$$
\delta \M(x) = x^3 g(x),~~ g(0) = 1,~~g'(0) = 0.\eqno(4.6)
$$
The boundary conditions at the outer boundary
are determined by the fact that the mass is conserved,
$$
\delta M = \delta\M(X) = 0.
$$
The equations then imply that the zero is of order $n$. With these
boundary conditions (4.5) becomes a well defined Sturm-Liouville problem
with an essentially self adjoint, second order differential operator.

Numerical calculations with the help of Mathematica are not difficult in the
case of integer values of $n$.
   It is found that, when $n= 2$ and for $n =3$,
$\delta\M(X)$ is positive in the whole range, for all negative values of
$\nu^2$ and for positive values below  a limit $\nu_0^2 $ that is about .06
for $n = 2$ and compatible with 0 for $n = 3$. The latter is the first,
nodeless solution of a sequence of solutions that we have not determined in
detail.  The function falls to zero at the surface,  where
  there is an $n$th order zero. Above this lowest value of
$\nu^2$ is a discrete set of other values of
$\nu^2$ at which the boundary condition is satisfied.

At the special value $n = 3$ the `ground state', the
lowest value of $\nu^2$, has approached very close to zero.

Polytropes with $n = 4$ are unstable.
We have searched for harmonic solutions with negative values
of $\nu^2$. The value $n = 4$ was chosen
because it is the only integer in the interesting range, and because
Mathematica is much more managable in this case.  (Accuracy is lost when
non integral powers of negative numbers appear at the end point.)
There seems to be a discrete, decaying nodeless mode with
$\nu^2 = -.015796$, but a bifurcation at this point in
parameter space makes the conclusion uncertain. We carried the calculation
to 15 significant figures in $\nu^2$ but solutions do not
converge towards a function that vanishes at the surface.
To overcome this difficulty we reformulated the problem in terms of the
variational calculus. The ``solution" found for $\nu^2 = -.015796$,
truncated near both ends, was used as a trial function, to show
conclusively that the spectrum of $\nu^2$ extends this far.
Among many papers on this topic we mention Cowling (1936) and Ledoux (1941).

\b\b

\ce{\bf V. The case $n = 3$}

This case  marks the boundary between stable and
unstable polytropes. The equations are conformally invariant
and a time independent solution is  found by an infinitesimal conformal
(homology) transformation,
$$
\delta f = rf' + f.\eqno(5.1)
$$
This does not represent an instability, but a ``flat direction", a
perturbation from which the system does not spring back, nor does it run
away. There must also be a second solution, linear in $t$, of the form
$$
\delta f = t(rf )',~~\delta\rho = t(r\rho ' + 3\rho ).
$$
The equation of continuity  becomes
$
   r\rho ' + 3\rho   + v\rho' + r^{-2}(r^2 v)'\rho   = 0,
$
whence $ v = -r$.

This linear perturbation is the first order approximation to the exact
solution found by
  Goldreich and Weber (1980), of the form
$$
f(r,t) = {1\over a(t)}\tilde f(x),~~ x = r/a(t).
$$
The continuity equation is solved by $v = \dot ax$; thus
$ \Phi = -(\dot a/a)(r^2/2) $, and
$$
\dot \phi - \vec v\,^2/2 = -a\dot a x^2/2 = cT + \phi.
$$
This leads to
$$
\tilde\phi := a(t)\phi \propto \tilde f + \kappa a^2\ddot a x^2/6,~~ \kappa
= 3k^{1/3}/c,
$$
  and Poisson's equation becomes
$$
\tilde f^3 +  {1\over x^2}(x^2\tilde f')' = {-\kappa\over x^2}a^2\ddot a
x^2/6 = - \kappa a^2\ddot a = \lambda,~~{\rm constant}.\eqno(5.2)
$$

There is a first integral,
$$
{\kappa\over 2} \dot a^2 - \lambda/a = C,~~{\rm constant}.
$$
Rescaling of $t$ and $a$ reduces this to one of three cases
$$
\dot a = \sqrt{1+1/a},~~ \dot a = \sqrt{1-1/a},~~ \dot a = 1/\sqrt a,
$$
but only the first is compatible with analyticity at $ t = 0$, thus
$$
t = \sqrt a\sqrt{1+a} - {\rm arcsinh}\sqrt a.
$$
Setting $ a = 1+b$ we find
$$
t = \sqrt{1/2}(b-b^2/2) + o(b^3)
$$
The factor $a(t)$ is  zero at a finite, negative value of $t$ and
increases monotoneously to infinity, passing through 1 at $t = 0$.
We can of course reverse the direction of flow of $t$ to get collapse in
the finite future.

Eq.(4.5) was solved numerically (Goldreich and Weber, 1980). The solution
is similar to the solution of Emden's equation, just prolonged a little at
the outer end, so long as $0<\lambda< .00654376$. For larger values of
$\lambda $ the distribution does not reach zero and increases for large $r$.
For simillar studies of collapsing, isothermal spheres see Hunter (1977) and
references therein.

It is confirmed, therefore, that the polytrope with $n = 3$ is not stable.
Suitably perturbed, the star may expand or collapse, until the higher or
lower density causes a change in the equation of state. Among may papers on
collapse we may mention Arnett (1977) and Van
Riper (1978).

\b\b

\no{\steptwo References}

\no Arnett, W.D., ``Neutrino trapping during gravitational
collapse of stars", 

Astrophys.J. {\bf 218}, 815-833 (1977).

\no Bernoulli, D., ~ Argentorat, 1738.

\no Boltzmann, L., Wissenschaftlidhe Abhandlungen, Hasenoehrl, Leipzig 1909.

\no Chandrasekhar, S., {\it An Introduction to Stellar Structure}, U.
Chicago Press 1938.

\no Collins, G.W.II,  "The virial Theorem in Stellar Astrophysics",

www.scribd.com/doc/7573482/ 

\no Cowling, T.G., M.N.R.A.S. {\bf 101}, 367 (1941).

\no Eddington, A.S., {\it The internal constitution of stars}, Dover, N.Y.
1959

\no Emden,  {\it Gaskugeln}, Teubner 1907.

 \no Fetter, A.L. and Walecka, J.D., {\it Theoretical Mechanics of Particles
and Continua}, McGraw-Hill, NY,1980.


\no Fronsdal, C., `` Ideal Stars and General Relativity", Gen.Rel.Grav.
(2007)

\no Fronsdal, C., ``Heat and Gravitation. I. The Action Principle",
arXiv:0812.4990.


\no Goldreich, P.  and Weber, S.V., ``Homologously collapsing stellar
models", 

Astrophys.J. {\bf 238}, 991-997 (1980). 

\no Hunter, C., ``The collapse of unstable isothermal stars", Astrophys. J.
{218}, 834-845 (1977).




255-260.


\no Kippenhahn, R.   and Weigert, A., ``Stellar Structure and Evolution",

Springer-Verlag, Berlin 1990.

\no Ledoux, P., Astrophys.J. {\bf 102}, 143 (1945).
 
\no Poisson, S.D., {\it Th\'eorie mathÈmatique de la chaleur},  1835.

\no Ritter, A., Wiedemann Annalen {\bf 11} 332 (1880). One of a series of
papers in Wiedemann 

Annalen, now Annalen der Physik.
 For a list see Chandrasekhar (1938). 
The volumes 

5-20 in
Wiedemann
Annalen appear as the volumes 241-256 in Annalen 
der Physik.

\no Thomson, W., Lord Kelvin, On Homer Lane's problem of a spherical
gaseous nebula,

Nature {\bf 75} 232-235 (1907).

\no Thomson, W., Lord Kelvin, On the convective equilibrium of temperature in
the

atmosphere, Manchester Phil.Soc. {\bf 2} , 170-176 (1862).

\no Tolman, R.C., {\it Relativity, Thermodynamics and Cosmology},
Clarendon, Oxford 1934.

\no Tolman, R.C., The electromotive force produced in solutions by
centrifugal action,

Phys.Chem. MIT, {\bf 59}, 121-147 (1910).

\no Van Riper, K.A., Astrophys.J. {\bf 221}, 304 (1978).

\no Weinberg, S., {\it Gravitation and Cosmology,...},    Wiley, 1972\ve

\end

\no Rees, M.F., Effects of very long wavelength primordial gravitational
radiation,

  Mon.Not.Astr.Soc. {\bf 154} 187-195 (1971).

\no Putterman, S. and Uhlenbeck, G.E., Thermodynamic equilibrium of Rotating
superfluids,

Phys. Fluids, {\bf 12}, 2229-2236 (1969).

\no  Davidson, R.D., {\it Theory of Non-Neutral Plasmas}, Addison-Wesley
1990.

\no Chandrasekhar, S. and Henrich, L.S., Stellar models with isothermal
cores,

Astrophys. J. {94}, 525-536 (1941).

\no Pipard, A.B.,  {\it Elements of Classical Thermodynamics}, Camb. U.
Press 1966.

\end

\b\b

\ce{\bf II.7. Virial theorem}

Both (2.14) and (2.16) are usually derived from considerations outside
the proper domain of thermodynamics. We  prefer
an axiomatic foundation of thermodynamics that is complete in the
sense that it does not need other input. As an  example
let us discuss the use of the virial theorem to make certain
predictions concerning stability.

The virial theorem was introduced into the present context by Kelvin (. It
is based on the scaling properties of the hamiltonian of a system of
particles. If
$H = K+V$, kinetic energy plus potential energy, then the lagrangian is
$K-V$ and the equations of motion imply that, up to a time derivative,
$$
\sum_im_i\dot q_i^2 = 2K = -\sum q_i\p_iV.
$$
In the case examined by Kelvin the potential is homogeneous of degree -1,
so that, \underbar{in the case of periodic motion}, when average is taken
over a period, $V = 2K$.  According to
  Chandrasekhar
  (1938) (pp. 49-51), who also quotes Poincar\'e, the internal energy
is the kinetic energy associated with  the
microscopic motion of the molecules. It is assumed, usually without
discussion, that the presence of gravitational forces do not affect the
internal energy, and that the total energy is obtained by simply adding the
gravitational potential energy to it. In the present approach there is no
place for this argument, the hamiltonian is the energy and there is only one
energy.

There is; however, a virial theorem associated with a lagrangian of
the type (2.7), that we abbreviate as
$$
{\cal L} = \rho(\dot \Phi - \vec v^2/2) - \hat V.
$$
(The potential $\hat V$ includes the gravitational field.) Variation of
$\Phi$ and of $\rho$ give the equations of motion
$$
\dot \Phi = \vec v^2/2+(d\hat V/d\rho),~~ \dot\rho = -{\rm div}(\rho
\vec v),
$$
which implies that
$$
\int d^3x {d\over dt}(\rho\Phi) = \int d^3x\Big(\rho{d\hat V\over d\rho} -
\rho\vec v\,^2/2\Big).
$$
If the system goes through a cycle then the average of this
quantity over the cycle is zero,
$$
<\int dx~\rho\vec v\,^2/2>~ = ~<\int dx~ \rho{d\hat V\over
d\rho}>.\eqno(2.17)
$$
  In the case of (2.10) we obtain, when $n = 3$,
$$
<\int dx~\rho\vec v\,^2/2>~ = ~<\int dx ~ \Big(\rho(\phi-\lambda) +
4\R T\rho +{4a\over 3}T^4\Big)>.\eqno(2.18)
$$
  With Eq.(2.13) this simplifies to
$$
<\int dx~\rho\vec v\,^2/2>~ = ~<\int dx ~ \Big(\rho(\phi-\lambda) +
\R(1+\log k)\rho T\Big)>.\eqno(2.19)
$$

  {\it This result, like classical virial theorems, applies
exclusively to the case of periodic motion.}

In the special case $\vec v = 0$ Eq.(2.19) is a direct consequence of the
equations of motion. Such relations, that do not depend on the periodicity
of the motion, are not true virial theorems.

\ve

\b\ve
\ve

\ve

\end

\b\b

\no{\steptwo V. General Relativity}

    \ce {\bf V.1. Lorentz invariance}

The limitation to small velocities, small compared to the velocity of
light, is justified almost always, with the sole exception of the photon
gas. We shall now modify our treatment of the non relativistic gas of
massive particles to make it consistent with relativistic invariance.

We need a 4-dimensional velocity and an associated
velocity potential,
$$
v_\mu = \p_\mu \psi  =:\psi_\mu,~~\mu = 0,1,2,3,
$$
where $\psi$ is a scalar field. There is only one reasonable lagrangian
(Fronsdal 2007),
$$
{\cal L} = {\rho\over 2}
\big(g^{\mu\nu}\psi_{,\mu}\psi_{,\nu} - c^2) - V[\rho].\eqno(5.1)
$$
The metric is the Lorentzian $g =~ $diag$ ~(c^{-2},-1,-1,-1)$.
In the case of velocities small compared to $c$ we set
$$
\psi = c^2t +  \Phi
$$
and find to order $o(c^{-2})$  the non relativistic lagrangian (1.10).
Henceforth   $c = 1$.

We easily allow for a dynamical gravitational field by
generalizing the measure,
$$
A = \int dt d^3x \sqrt{-g}\,{\cal L}.
$$
In a weak, terrestrial gravitational field the usual approximation for the
metric is\break
$
g =~ {\rm diag.} ~ (1-2gz,-1,-1,-1),
$
which leads to (2.10).

The concept of energy (density) is all-important in thermodynamics
    and in relativistic field theories
but ill defined in General Relativity.
    However, as long as we
limit our attention to time independent configurations, we expect to be
on relatively safe grounds when we identify the energy density with the
time-time component of the energy-momentum tensor,
$$
T_{\mu\nu} =   \rho \psi_{,\mu}\psi_{,\nu} -
g_{\mu\nu}{\cal L} .\eqno(5.2)
$$
In the non relativistic limit $T_{00}$ is our hamiltonian
    augmented with the rest mass.

The Euler-Lagrange equations include  the conservation law
$$
\p_\mu J^\mu = 0,~~ J^\mu := \sqrt{-g}g^{\mu\nu}\psi_{,\nu}.
$$
The integral $
\int\sqrt{-g}\rho d^3x$ is a constant of the motion (for appropriate
boundary conditions) and can be interpreted as mass. This is an
essential improvement over the traditional treatment. A conserved current
also permits an application to a non neutral plasma (Fronsdal 2007).
The (conserved) mass plays a central role
in fixing the boundary conditions in the non relativistic theory;
to retain this feature in the relativistic extension is natural.

\b

\ce {\bf V.2. Polytropic star with radiation}

Here  we propose to try out the lagrangian (2.7) or its relativistic 
version for
the mixture of an ideal gas with the photon gas. In the case that the
radiation pressure is relatively unimportant there is nothing new in this,
and in the  special case that $n = 3$ the theory is identical with that of
Eddington.

In the relativistic case the action principle offers advantages even in
this particular case. Clarification of the role of mass, which is
confused or at least confusing in the traditional treatment, is an
important part of it. Another advantage is the relative ease with which one
may proceed to study mixtures.

Variation of the action with respect to the
temperature gives the relation  (II.10) that shows a departure from the
polytropic relation $\rho = kT^n$ when $n \neq 3$. (If this last relation is
accepted, in lieu of (II.10),
  then from this point on the equations of motion are the same
as with other methods.) The relation between Eddington's parameter
$\beta$ and $k,n$ is
$$
{1\over \beta}  =  p_{\rm tot}/ p_{\rm gas}  =   1 +
{a\over 3\R k};
$$
It is constant only when $n = 3$.
In the relativistic theory, the same relations hold; Eq.(II.10) remains
valid. The equation that determines the temperature is transcendental;
the first approximation is   $\log k/k_0 = 3$.

Applications to real stars should await the incorporation of heat flow, not
important in the case of an isolated atmosphere and of secondary
importance in the case of the earthly atmosphere, but perhaps vital for a full
understanding of the physics.

  \no{\steptwo VI. Conclusions}

\ce{\bf VI.1. On variational principles}

Variational principles have a very high reputation in most branches of
physics; they even occupy a central position in classical thermodynamics,
see for example the authoritative treatment by Callen (1960).
An action is available for the study of laminar flows in hydrodynamics, see
e.g. Fetter and Walecka (1960), though it does not seem to have been much used.
Without the restriction to laminar flows it remains possible to
formulate an action principle (Taub 1954, Bardeen  1970, Schutz  1970), but
the proliferation of velocity potentials is confusing and
no application is known to us. Recently, variational principles have been
invoked in special situations that arise in gravitation.

In this paper we rely on an action principle formulation of the full set of
laws that govern an ideal gas, in the presence of gravity and radiation.
To keep it simple we have restricted our attention to laminar,
hydrodynamical  flows.

It was shown that there is an action that incorporates both of Poisson's
laws as variational equations, the temperature field being treated as any
other dynamical variable. The idea of
varying the action with respect to the temperature is much in the classical
tradition. The variational equations of motion are exactly the classical
relations if radiation is neglected, or if $n = 3$.

The first encouraging result comes with the realization that
the hamiltonian gives the correct expression for the internal
energy and the pressure, including the contributions of
radiation, under the circumstances that are considered in classical
thermodynamics; that is, in equilibrium and in the absence of gravitation. This
is an indication that the theory is mathematically complete, requiring no
additional input from the underlying microscopic interpretation. This 
conclusion
is reinforced by an internal derivation of a virial theorem.

Into this framework the inclusion of a gravitational field is natural.
Inevitably, it leads to pressure gradients and thus also temperature gradients.
If other considerations, including the heat equation, are put aside, then the
theory, as it stands, predicts the persistence of a temperature 
gradient in an
isolated system at equilibrium. The existence of a 
temperature gradient in an
isolated thermodynamical system is 
anathema to tradition, and further work is
required to find the way 
to avoid it, or to live with it. Physical
considerations indicate 
that the answer is to be found in the phenomenon of
convection. The 
theory in the present form can be applied when convection is 
not
important.

A secondary but satisfying result of this work has 
been the application of
the action principle to the study of the 
energy concept. Without a well
defined hamiltonian it is quite 
impossible to attach an operative meaning to
any expression for the 
value of the energy; it is always
defined up to an additive constant, 
independently for each solution  of the
equations of motion. With a 
hamiltonian at our disposal we are  in a
position to give voice to 
our misgivings concerning the way that ``energy"
has been invoked in 
some branches of physics over a period of over 100 years.
Though we 
conclude that past demonstrations of instabilities of polytropes
are 
inconclusive, we do not suggest that the results are wrong. It is of 
course
agreed that $n = 3$ represents an important bifurcation point.

We have insisted on the role played by the mass in
fixing the 
boundary conditions, verified for 3 different
atmospheres. The 
existence of a conserved current and the associated
constant of the 
motion is especially important in the context
of General Relativity 
where the absence of this concept casts a shadow of
doubt on the 
choice of boundary conditions (Fronsdal 2008). Indeed it is
strange 
that the equation of continuity, a major pillar of 
nonrelativistic
hydrodynamics, has been abandoned without protest in 
the popular
relativistic extension. See Kippenhahn and Weigert 
(1990), pages 12-13.
\b
The interaction of the ideal gas with 
electromagnetic fields has been
discussed in a provisional manner. 
The transfer of
entropy between the two gases is in accord with the 
usual treatment of each
system separately.
 
 \b 

\ce{\bf VI.2. 
Suggestions} 

(1) It is suggested that observation of the diurnal 
and seasonal
variations of the equation of state of the troposphere 
may lead to a better
understanding of the role of radiation in our 
atmosphere. The centrifuge
may also be a practical source of 
enlightenment. We understand that modern
centrifuges are capable of 
producing accelerations of up to 10$^6 g$.
Any positive result for 
the temperature gradient in an isolated gas would
certainly have 
important theoretical consequences.

(2)   We suggest the use of 
the
lagrangian (2.7), or its relativistic extension, with
$T$ treated 
as an independent dynamical variable and
   $n' = n$. Variation with 
respect to $T$
yields  the adiabatic relations between $\rho$ 
and
$T$, so long as the pressure of radiation is negligible, but 
for
higher temperatures, when radiation becomes important, the effect 
is to increase
the effective value of $n'$ towards the ultimate limit 
3, regardless of the
adiabatic index $n$ of the gas. See in this 
connection the discussion by
Cox and Giuli (1968),  page 
271.\break
In the case that $n = 3$ there is Eddington's treatment of 
the
mixture of an ideal gas with the photon gas.  But most gas 
spheres have a
polytropic index somewhat less than 3 and in this case 

the ratio $\beta = p_{\rm gas}/p_{\rm tot}$ may not be constant 
throughout
the star. The lagrangian (2.7), with $n$ identified with 
the 
adiabatic index of the gas, gives all the equations that are 
used to
describe atmospheres, so long as radiation is 
insignificant.
With greater radiative pressure the polytropic index 
of the atmosphere is
affected. It is not quite constant, but  nearly 
so, and it approaches
the upper limit 3 when the radiation pressure 
becomes dominant. Eddington's
treatment was indicated because he used 
Tolman's approach to relativistic
thermodynamics, where there is room 
for only one density and only one
pressure.   Of course, all kinds of 
mixtures have been studied, but the
equations that govern them do not supplement Tolman's gravitational concepts in
a satisfactory  manner, in our opinion. Be that as it may, it is 
patent that the
approximation
$\beta $ = constant, in the works of Eddington and Chandrasekhar, is a device
designed  to avoid dealing with two independent gases.

\b\b
\no{\steptwo Acknowledgements}

I thank R.J. Finkelstein, R.W. Huff and P. Ventrinelli for discussions.

\b\b

\no{\steptwo References}

\no Arnett, W.D., Ap.J. Suppl., {\bf 35}, 145 (1977). Ap. J. {bf 218}, 815
(1977).

\no Bardeen, J.M., A variational principle for rotating stars
in General Relativity,

  Astrophys. J. {162}, 7 (1970).

\no Bernoulli, D., ~ Argentorat, 1738.

\no Boltzmann, L., Wissenschaftlidhe Abhandlungen, Hasenoehrl, Leipzig 1909.

\no Callen, H.B., {\it Thermodynamics}, John Wiley N.Y. 1960. 

\no Carnot, S., quoted by Emden (1907).

\no Castor, J., {\it Radiation Hydrodynamics}, Cambridge U. press,  2004.

\no Chandrasekhar, S., {\it An Introduction to Stellar Structure}, U.
Chicago Press 1938.

\no Cheng, A.F., Unsteady hydrodynamics of spherical gravitational
collapse,

Astrophys. J., {\bf 221}, 320-326 (1978).

\no Cox, J.P. and Giuli, R.T., Principles of stellar structure, Gordon and
Breach, 1968.

\no Cowling, T.G., Monthly Notices Astr. Soc. {\bf 96}, 42 (1936).

\no De Groot, S.R., {\it An Introduction to Modern Thermodynamical
Principles}, Oxford U.

Press, 1937. 

\no Emden,  {\it Gaskugeln}, Teubner 1907.

\no Euler, H.,   {\it 
\"Uber die Streuung von Licht an Licht nach der
Diracschen 
Theorie},

 Ann.Phys.  {\bf 26} 398-?  (1936).

\no Fetter, A.L. and 
Walecka, J.D., {\it Theoretical Mechanics of Particles
and 
Continua},
 
\no Finkelstein, R.J., {\it Thermodynamics and 
statistical physics}, W.H.
Freeman 1969.

\no Fronsdal,  C.,  Ideal 
Stars and General Relativity,  

Gen.Rel.Grav. {\bf 39} 1971-2000 
(2007), gr-qc/0606027.

\no Fronsdal, C., Reissner-Nordstrom and 
charged polytropes,

Lett.Math.Phys. {\bf 82}, 255-273 (2007).  

\no 
Fronsdal, C., Stability of polytropes, Phys. Rev. D. (to appear), 
arXiv
0705.0774 [gr-cc].

\no Goldreich, P. and Weber, S.V., 
Homologously collapsing
stellar cores, 

Astrophys.J. {\bf 238} 
991-997 (1980).

\no Graeff, R.W., Viewing the controversy 
Loschmidt-Boltzmann/Maxwell through

macroscopic measurements of the 
temperature gradients in vertical columns of
water, 

 preprint 
(2007).  

\no Holman, J.P. {\it Thermodynamics}, McGraw-Hill, N.Y. 
1969.

\no Hunter, C., Collapse of unstable isothermal apheres, 
Astrophys. J.,
{\bf 218} 834-845 (1977).  

\no Karplus, R. and 
Neuman, M.,
Non-linear Interactions between Electromagnetic Fields,

Phys.Rev. {\bf 80} 380-385 (1950).

\no Kelvin, Thomson, W., 
Collected Mathematical and Physical papers, Vol. 5,
 232-235. 

\ve

\no Kelvin, Thomson, W., Collected Mathematical and Physical 
papers, Vol. 3,
255-260. 

Cambridge U. Press 1911.
  
\no 
Kippenhahn, R. and Weigert, A, ``Stellar Structure and
Evolution", 
Springer-Verlag 1990.

equilibrium, governed by equations

gradient".

\no Lane, H.J.,   On the Theoretical Temperature of the 
Sun, under the
Hypothesis of a gaseous 

Mass maintaining its Volume 
by its internal Heat,
and depending on the laws of gases 

as known 
to terrestrial Experiment, Amer.J.Sci.Arts, Series 2, {\bf 4}, 57- 

(1870).

\no Ledoux, P., On the vibrational stability of gaseous 
stars, Astrophys.
J., {\bf 94}, 537-548 (1941).

\no Loschmidt, L., 
Sitzungsb. Math.-Naturw.
Klasse Kais. Akad. Wissen. {\bf 73.2} 135 
(1876).  

 \no Maxwell, J.C., The London, Edinburgh and Dublin 
Philosophical Magazine
{\bf35} 215 (1868).

\no Mazur, P. and 
Mottola, E., Gravitational Vacuum Condensate Stars,

gr-qc/0407075

\no McKenna, J. and Platzman,  P.M., 
Nonlinear
Interaction of Light in Vacuum,

  Phys. Rev. {\bf 129} 
2354-2360 (1962).

\no Milne,E.A., Statistical Equilibrium in 
relation to the Photoelectric
Effect and its 

Application to the 
Determination of Absorption Coefficients,
Phi.Mag. {\bf 47}, 209 
(1924).

\no Milne, E.A., Polytropic equilibrium 1.  the effect of 
configurations
 under given external 

?Handbuch Astrophys. {\bf 3} 
(1930). Pages 204-222.


\no M\"uller, I., {\it A History 
of Thermodynamics}, Springer, Berlin 2007. 

\no Panofsky W.K.H.  and 
Philips, M., {\it Classical Electricity and
Magnetism},

Addison-Wesley, Reading Mass.  1962.

\no Poisson, S.D., {\it 
Th\'eorie mathÈmatique de la chaleur},  1835.

\no Ritter, A., A 
series of papers in Wiedemann Annalen, now Annalen der
Physik,

 For 
a list see Chandrasekhar (1938). The volumes 5-20 in
Wiedemann

Annalen appear as the volumes 241-256 in Annalen der Physik.

\no 
Rosseland, S., Oslo Pub., No. 1, 1931.

\no Rosseland, S. {\it 
Theoretical Astrophysics}, Oxford U. Press 1936.
 
\no Saha,  M.N. 
and Srivastava, B.N., {\it A treatise on heat}, The Indian
Press, 
1935. 

\no Schutz, B.F. Jr., Perfect fluids in General Relativity: 
Velocity
potentials and a  variational 

principle, Phys.Rev.D {\bf 
2}, 2762-2771 (1970). 

\no Schwarzschild, K., Ueber das 
Gleighgewicht der Sonnenatmosph\"are,

G\"ottinger Nachrichten, 
41-53 (1906).

\no Stanyukovich, K.P., {\it Unsteady motion of 
continuous media},
Pergamon Press N. Y. 1960.
 
{\it Thermodynamics and Statistical Mechanics}, Lectures
Theoretical Physics,  

QC311 S69tE.

\no Taub, A.H., General relativistic variational 
principle for
perfect fluids, 

Phys.Rev. {\bf 94}, 1468 (1954).

\no 
Thomson, W., Lord Kelvin, On Homer Lane's problem of a 
spherical
gaseous nebula, 

Nature {\bf 75} 232-235 (1907).

\no 
Thomson, W., Lord Kelvin, On the convective equilibrium of 
temperature in
the 

atmosphere, Manchester Phil.Soc. {\bf 2} , 
170-176 (1862). 

\no Tolman, R.C., {\it Relativity, Thermodynamics 
and Cosmology},
Clarendon, Oxford 1934.

\no Tolman, R.C., The 
electromotive force produced in solutions by
centrifugal action,

Phys.Chem. MIT, {\bf 59}, 121-147 (1910). 

\no Van Riper, K.A., 
The hydrodynamics of stellar collapse, Astrophys. J.
{\bf 221} 
304-319 (1978).
  
\no Vanderslice et al, {\it Thermodynamics}, 
Prentice-Hall 1966.

\no Waldram, J.R., {\it The theory of 
electrodynamics}, Cambridge U. Press
1985. 
\ve

\end

\no Rees, M.F., Effects of very long wavelength primordial gravitational
radiation,

Mon.Not.Astr.Soc. {\bf 154} 187-195 (1971).

\no Putterman, S. and 
Uhlenbeck, G.E., Thermodynamic equilibrium of Rotating
superfluids,

Phys. Fluids, {\bf 12}, 2229-2236 (1969).

\no  Davidson, R.D., 
{\it Theory of Non-Neutral Plasmas}, Addison-Wesley
1990.

\no 
Chandrasekhar, S. and Henrich, L.S., Stellar models with 
isothermal
cores, 

Astrophys. J. {94}, 525-536 (1941).

\no 
Pipard, A.B.,  {\it Elements of Classical Thermodynamics}, Camb. 
U.
Press 1966.

\end
\b\b 

\no{\steptwo VI. 
Conclusions}

\ce{\bf VI.1. On variational principles}

The 
principal reasons for preferring an action principle formulation of 
thermodynamics were stressed in the introduction. Here we add some 
additional comments.

Variational principles have a very high 
reputation in most branches of
physics; they even occupy a central 
position in classical thermodynamics,
see for example the 
authoritative treatment by Callen (1960).
An action is available for 
the study of laminar flows in hydrodynamics, see
e.g. Fetter and 
Walecka (1960), though it does not seem to have been much 
used.
Without the restriction to laminar flows it remains possible 
to
formulate an action principle (Taub 1954, Bardeen  1970, Schutz 
1970), but
the proliferation of velocity potentials is confusing 
and
no application is known to us. Recently, variational principles 
have been
invoked in special situations that arise in 
gravitation.

In this paper we rely on an action principle 
formulation of the full set of
laws that govern an ideal gas, in the 
presence of gravity and radiation.
To keep it simple we have 
restricted our attention to laminar,
hydrodynamical  flows.  

It was 
shown that there is an action that incorporates both of 
Poisson's
laws as variational equations, the temperature field being 
treated as any
other dynamical variable. The idea of
varying the 
action with respect to the temperature is much in the 
classical
tradition. The variational equations of motion are exactly 
the classical
relations if radiation is neglected, or if $n = 3$.

The first encouraging result comes with the realization that
the 
hamiltonian gives the correct expression for the internal
energy and 
the pressure, including the contributions of 
radiation, under the 
circumstances that are considered in classical
thermodynamics; that 
is, in equilibrium and in the absence of gravitation. This
is an 
indication that the theory is mathematically complete, requiring 
no
additional input from the underlying microscopic interpretation. 
This conclusion
is reinforced by an internal derivation of a virial 
theorem.

Into this framework the inclusion of a gravitational field 
is natural. 
Inevitably, it leads to pressure gradients and thus also 
temperature gradients.
If other considerations, including the heat 
equation, are put aside, then the
theory, as it stands, predicts the 
persistence of a temperature gradient in an
isolated system at 
equilibrium. The existence of a temperature gradient in an
isolated 
thermodynamical system is anathema to tradition, and further work 
is
required to find the way to avoid it, or to live with it. 
Physical
considerations indicate that the answer is to be found in 
the phenomenon of
convection. The theory in the present form can be 
applied when convection is not
important.

A secondary but 
satisfying result of this work has been the application of
the action 
principle to the study of the energy concept. Without a well
defined 
hamiltonian it is quite impossible to attach an operative meaning 
to
any expression for the value of the energy; it is always
defined 
up to an additive constant, independently for each solution  of 
the
equations of motion. With a hamiltonian at our disposal we are 
in a
position to give voice to our misgivings concerning the way that 
``energy"
has been invoked in some branches of physics over a period 
of over 100 years.
Though we conclude that past demonstrations of 
instabilities of polytropes
are inconclusive, we do not suggest that 
the results are wrong. It is of course
agreed that $n = 3$ represents 
an important bifurcation point. 
 
We have insisted on the role 
played by the mass in
fixing the boundary conditions, verified for 3 
different
atmospheres. The existence of a conserved current and the 
associated
constant of the motion is especially important in the 
context
of General Relativity where the absence of this concept casts 
a shadow of
doubt on the choice of boundary conditions (Fronsdal 
2008). Indeed it is
strange that the equation of continuity, a major 
pillar of nonrelativistic
hydrodynamics, has been abandoned without 
protest in the popular
relativistic extension. See Kippenhahn and 
Weigert  (1990), pages 12-13.
\b
The interaction of the ideal gas 
with electromagnetic fields has been
discussed in a provisional 
manner. The transfer of
entropy between the two gases is in accord 
with the usual treatment of each
system separately.
 
 \b 

\ce{\bf 
VI.2. Suggestions} 

(1) It is suggested that observation of the 
diurnal and seasonal
variations of the equation of state of the 
troposphere may lead to a better
understanding of the role of 
radiation in our atmosphere. The centrifuge
may also be a practical 
source of enlightenment. We understand that modern
centrifuges are 
capable of producing accelerations of up to 10$^6 g$.
Any positive 
result for the temperature gradient in an isolated gas 
would
certainly have important theoretical consequences.

(2)   We 
suggest the use of the
lagrangian (2.7), or its relativistic 
extension, with
$T$ treated as an independent dynamical variable and
 
$n' = n$. Variation with respect to $T$
yields  the adiabatic 
relations between $\rho$ and
$T$, so long as the pressure of 
radiation is negligible, but for
higher temperatures, when radiation 
becomes important, the effect is to increase
the effective value of 
$n'$ towards the ultimate limit 3, regardless of the
adiabatic index 
$n$ of the gas. See in this connection the discussion by
Cox and 
Giuli (1968),  page 271.\break
In the case that $n = 3$ there is 
Eddington's treatment of the
mixture of an ideal gas with the photon 
gas.  But most gas spheres have a
polytropic index somewhat less than 
3 and in this case 
the ratio $\beta = p_{\rm gas}/p_{\rm tot}$ may 
not be constant throughout
the star. The lagrangian (2.7), with $n$ 
identified with the 
adiabatic index of the gas, gives all the 
equations that are used to
describe atmospheres, so long as radiation 
is insignificant.
With greater radiative pressure the polytropic 
index of the atmosphere is
affected. It is not quite constant, but 
nearly so, and it approaches
the upper limit 3 when the radiation 
pressure becomes dominant. Eddington's
treatment was indicated 
because he used Tolman's approach to relativistic
thermodynamics, 
where there is room for only one density and only one
pressure.   Of 
course, all kinds of mixtures have been studied, but the
equations that govern them do not supplement Tolman's gravitational concepts in
a satisfactory  manner, in our opinion. Be that as it may, it is 
patent that the
approximation
$\beta $ = constant, in the works of Eddington and Chandrasekhar, is a device
designed  to avoid dealing with two independent gases.

\b\b
\no{\steptwo Acknowledgements}

I thank R.J. Finkelstein, R.W. Huff and P. Ventrinelli for discussions.

\b\b

\no{\steptwo References}

\no Arnett, W.D., Ap.J. Suppl., {\bf 35}, 145 (1977). Ap. J. {bf 218}, 815
(1977).

\no Bardeen, J.M., A variational principle for rotating stars
in General Relativity,

  Astrophys. J. {162}, 7 (1970).

\no Bernoulli, D., ~ Argentorat, 1738.

\no Boltzmann, L., Wissenschaftlidhe Abhandlungen, Hasenoehrl, Leipzig 1909.

\no Callen, H.B., {\it Thermodynamics}, John Wiley N.Y. 1960. 

\no Carnot, S., quoted by Emden (1907).

\no Castor, J., {\it Radiation Hydrodynamics}, Cambridge U. press,  2004.

\no Chandrasekhar, S., {\it An Introduction to Stellar Structure}, U.
Chicago Press 1938.

\no Cheng, A.F., Unsteady hydrodynamics of spherical gravitational
collapse,

Astrophys. J., {\bf 221}, 320-326 (1978).

\no Cox, J.P. and Giuli, R.T., Principles of stellar structure, Gordon and
Breach, 1968.


\no De Groot, S.R., {\it An Introduction to Modern Thermodynamical
Principles}, Oxford U.

Press, 1937. 

\no Emden,  {\it Gaskugeln}, Teubner 1907.

\no Euler, H.,   {\it 
\"Uber die Streuung von Licht an Licht nach der
Diracschen 
Theorie},

 Ann.Phys.  {\bf 26} 398-?  (1936).

\no Fetter, A.L. and 
Walecka, J.D., {\it Theoretical Mechanics of Particles
and 
Continua},
 
\no Finkelstein, R.J., {\it Thermodynamics and 
statistical physics}, W.H.
Freeman 1969.

\no Fronsdal,  C.,  Ideal 
Stars and General Relativity,  

Gen.Rel.Grav. {\bf 39} 1971-2000 
(2007), gr-qc/0606027.

\no Fronsdal, C., Reissner-Nordstrom and 
charged polytropes,

Lett.Math.Phys. {\bf 82}, 255-273 (2007).  

\no 
Fronsdal, C., Stability of polytropes, Phys. Rev. D. (to appear), 
arXiv
0705.0774 [gr-cc].

\no Goldreich, P. and Weber, S.V., 
Homologously collapsing
stellar cores, 

Astrophys.J. {\bf 238} 
991-997 (1980).

\no Graeff, R.W., Viewing the controversy 
Loschmidt-Boltzmann/Maxwell through

macroscopic measurements of the 
temperature gradients in vertical columns of
water, 

 preprint 
(2007).  

\no Holman, J.P. {\it Thermodynamics}, McGraw-Hill, N.Y. 
1969.

\no Hunter, C., Collapse of unstable isothermal apheres, 
Astrophys. J.,
{\bf 218} 834-845 (1977).  

\no Karplus, R. and 
Neuman, M.,
Non-linear Interactions between Electromagnetic Fields,

Phys.Rev. {\bf 80} 380-385 (1950).

\no Kelvin, Thomson, W., 
Collected Mathematical and Physical papers, Vol. 5,
 232-235. 

\ve

\no Kelvin, Thomson, W., Collected Mathematical and Physical 
papers, Vol. 3,
255-260. 

Cambridge U. Press 1911.
  
\no 
Kippenhahn, R. and Weigert, A, ``Stellar Structure and
Evolution", 
Springer-Verlag 1990.

equilibrium, governed by equations

gradient".

\no Lane, H.J.,   On the Theoretical Temperature of the 
Sun, under the
Hypothesis of a gaseous 

Mass maintaining its Volume 
by its internal Heat,
and depending on the laws of gases 

as known 
to terrestrial Experiment, Amer.J.Sci.Arts, Series 2, {\bf 4}, 57- 

(1870).

\no Ledoux, P., On the vibrational stability of gaseous 
stars, Astrophys.
J., {\bf 94}, 537-548 (1941).

\no Loschmidt, L., 
Sitzungsb. Math.-Naturw.
Klasse Kais. Akad. Wissen. {\bf 73.2} 135 
(1876).  

 \no Maxwell, J.C., The London, Edinburgh and Dublin 
Philosophical Magazine
{\bf35} 215 (1868).

\no Mazur, P. and 
Mottola, E., Gravitational Vacuum Condensate Stars,

gr-qc/0407075

\no McKenna, J. and Platzman,  P.M., 
Nonlinear
Interaction of Light in Vacuum,

  Phys. Rev. {\bf 129} 
2354-2360 (1962).

\no Milne,E.A., Statistical Equilibrium in 
relation to the Photoelectric
Effect and its 

Application to the 
Determination of Absorption Coefficients,
Phi.Mag. {\bf 47}, 209 
(1924).

\no Milne, E.A., Polytropic equilibrium 1.  the effect of 
configurations
 under given external 

?Handbuch Astrophys. {\bf 3} 
(1930). Pages 204-222.


\no M\"uller, I., {\it A History 
of Thermodynamics}, Springer, Berlin 2007. 

\no Panofsky W.K.H.  and 
Philips, M., {\it Classical Electricity and
Magnetism},

Addison-Wesley, Reading Mass.  1962.

\no Poisson, S.D., {\it 
Th\'eorie mathÈmatique de la chaleur},  1835.

\no Ritter, A., A 
series of papers in Wiedemann Annalen, now Annalen der
Physik,

 For 
a list see Chandrasekhar (1938). The volumes 5-20 in
Wiedemann

Annalen appear as the volumes 241-256 in Annalen der Physik.

\no 
Rosseland, S., Oslo Pub., No. 1, 1931.

\no Rosseland, S. {\it 
Theoretical Astrophysics}, Oxford U. Press 1936.
 
\no Saha,  M.N. 
and Srivastava, B.N., {\it A treatise on heat}, The Indian
Press, 
1935. 

\no Schutz, B.F. Jr., Perfect fluids in General Relativity: 
Velocity
potentials and a  variational 

principle, Phys.Rev.D {\bf 
2}, 2762-2771 (1970). 

\no Schwarzschild, K., Ueber das 
Gleighgewicht der Sonnenatmosph\"are,

G\"ottinger Nachrichten, 
41-53 (1906).

\no Stanyukovich, K.P., {\it Unsteady motion of 
continuous media},
Pergamon Press N. Y. 1960.



\no Taub, A.H., General relativistic variational principle for
perfect fluids,

Phys.Rev. {\bf 94}, 1468 (1954).

\no Thomson, W., Lord Kelvin, On Homer Lane's problem of a spherical
gaseous nebula,

Nature {\bf 75} 232-235 (1907).

\no Thomson, W., Lord Kelvin, On the convective equilibrium of temperature in
the

atmosphere, Manchester Phil.Soc. {\bf 2} , 170-176 (1862).

\no Tolman, R.C., {\it Relativity, Thermodynamics and Cosmology},
Clarendon, Oxford 1934.

\no Tolman, R.C., The electromotive force produced in solutions by
centrifugal action,

Phys.Chem. MIT, {\bf 59}, 121-147 (1910).

\no Van Riper, K.A., The hydrodynamics of stellar collapse, Astrophys. J.
{\bf 221} 304-319 (1978).

\no Vanderslice et al, {\it Thermodynamics}, Prentice-Hall 1966.

\no Waldram, J.R., {\it The theory of electrodynamics}, Cambridge U. Press
1985.
\ve

\end

\no Rees, M.F., Effects of very long wavelength primordial gravitational
radiation,

  Mon.Not.Astr.Soc. {\bf 154} 187-195 (1971).

\no Putterman, S. and Uhlenbeck, G.E., Thermodynamic equilibrium of Rotating
superfluids,

Phys. Fluids, {\bf 12}, 2229-2236 (1969).

\no  Davidson, R.D., {\it Theory of Non-Neutral Plasmas}, Addison-Wesley
1990.

\no Chandrasekhar, S. and Henrich, L.S., Stellar models with isothermal
cores,

Astrophys. J. {94}, 525-536 (1941).

\no Pipard, A.B.,  {\it Elements of Classical Thermodynamics}, Camb. U.
Press 1966.

The lagrangian (2.7) is thus   successful in accounting for the
properties of a polytropic atmosphere (not just the adiabatic atmosphere)
constituted by an ideal gas. If the gravitational force is included  it
predicts a temperature gradient that is verified experimentally.
Till now we have seen no explanation of the strange fact that
real atmospheres are polytropic. The usual interpretation is, we 
think, that the
temperature gradient is created by the incoming radiation.
Emden's remark (Emden 1907, page 320) on heat flow reminds us, in the first
place, of the physical reality of heat flow. Physics demands that it be
continuous. Since the gradient of the temperature is not zero at the lower
boundary heat is entering from below and the system is not isolated.

Let us accept that the approach to equilibrium is accompanied by heat flow from
hot to cold, at a rate that depends on the properties of the gas and on its
state,   through the heat equation. Our present lagrangian describes a
situation in which heat flow is absent from the equations of motion;
there are no terms involving derivatives of $T$,
temporal or spacial. The absence of
$\dot T$ is natural, since we are describing stationary states.
But the heat equation   (we suppose that it is valid although
it is not incorporated into our dynamical framework) then implies that $C\vec
\btd T$ is constant. Emden suggests that this is indeed the case, and that the
heat equation is satisfied. If that is so, then we understand that the only
way that the flow enters into the dynamics of the polytropic atmosphere is
through the boundary conditions.

{\it What we find unsatisfactory, however, is the fact that there is no clear
relationship betwee the adiabatic lagrangian and the isolated, isothermal gas.}
Ideally, a parameter should be present that would allow to reduce the intensity
of radiation to zero, ending with the isolated, isothermal atmosphere in the
limit. The parameter $k_0$ determines the temperature gradient,
but  the polytropic index is unaffected by a change of $k_0$.

There is a concept of convection driven by the radiation,
though it is difficult to justify it in the case that
$\gamma'>\gamma$, when the atmosphere is stable to convection. But convection
must cease in the final approach of
an isolated gas to equilibrium. The mathematical difficulty may be 
due, in part,
to the fact that the mass flow is not taken into account.

The simple answer to this is that the isolated atmosphere remains adiabatic!
That is not in contradiction with the heat equation, but it would imply a
constant heat flow that is difficult to accept. Yet, this would provide a neat
explanation of why the polytropic index takes a value that depends strongly on
the properties of the gas but very weakly on the radiation.

In Section II.3 we shall try to say something about interactions with an
electric field. The problem of incorporating heat flow into the lagrangian is
left for the future.

\ce{\steptwo III. Nonstationary configurations}

\ce{\bf 3.1. }

In an isolated system (now we know that atmospheres are not usually isolated),
a temperature gradient implies change, the variables are not constant in time.
Therefore, we need to introduce the gradient of $T$ into the action.
Let us try adding a term
$$
C[\rho,T](\vec\btd T)^2/2.
$$
It looks like the contribution of kinetic energy and implies that the heat flow
possesses inertia. This will affect the equation that comes from variation of
$T$ and none of the others. It will not imply that the temperature change
according to Fourier's law, or any temporal change at all. We need to introduce
$\dot T$ into the lagrangian.

This is difficult, for any term that is linear in $\dot T$, of the form
$f(T)\dot T$ is a time derivative and will make no contribution to the
Euler-Lagrange equations. The first order temporal; derivative $\dot 
T$ can only
appear in the variational equation if it appears in the lagrangian. A term like
$T\dot T$ makes no contribution to the Euler-Lagrange equations. Variation of
$T$ gives an equation that does not ressemble Fourier's heat 
equation. An example
of a term that gives the correct equation is
$$
S\dot T + C[\rho,T] \,\,\vec\btd S\cdot\vec\btd T .
$$
What makes this work is the fact that the additional variable $S$ 
does not occur
elsewhere in the lagrangian; indeed this may be the only way to get the heat
equation from the lagrangian. The only difficulty with this is that 
the physical
interpretation of the new field is unclear. Let us ignore that and work out the
properties of the theory obtained by adding this term to the lagrangian (2.10).

\b\b

\ce{\bf 3.3. The dielectric atmosphere}

Consider the lagrangian
$$
{\cal L} = \rho(\dot\Phi - \vec v^2/2 -gz + \lambda )
   -{\cal 
R}T\rho\log k +   f[T] + {c\rho\over T}\overline{F^2}/2.
 
\eqno(A.1)
$$
The factor  $c\rho /T$ has the correct dependence
 on 
density and temperature,  to be interpreted as permittivity.
The 
average $\overline{F^2}$ is not determined by the 
average
$\overline{F_{\mu\nu}}$ of the electromagnetic field 
strength.  Therefore,
it is risky to identify
$c\rho /T$ with the 
numerical value of the permittivity of air, which is
00054 for dry 
air at sea level, and we have nothing to say about the
observed 
electric field in the atmosphere.

The equations of motion 
are
$$
\lambda - gz -  \R T\big(1 + \log k)+{c\over 
T}\overline{F^2}/2  = 0,
\eqno(A.4)
$$
$$
\R \rho  (n-\log k)
+ 
{4a\over 3} T^3  - {c\rho\over T^2}\overline{F^2}/2= 
0,\eqno(A.5)
$$
If we neglect the radiation term we must have
$$
\R 
\rho  (n-\log k)
={c\rho\over T }\overline{F^2}/2T,\eqno(A.5)
$$
An 
increase of 10 percent in the sea level temperature
 leads to a 
decrease of $\log k$ of about  .286.
The accompanying change in the 
density is comparatively
 less important and we find that
the 
required change in $\overline{F^2}$ is
$$
\delta\overline{F^2} = 
.286
  {(2.87\times 10^6) \times10^{-3}\times 2\times 294\over 
5\times 10^{-4}}
=.965 \times 10^9.
$$
If this were due to a uniform 
electric field it would have a
 strength of  more than $3.1\times 
10^4$ in cgs units or  3 million
$volts/cm$.\footnote*{Verify.}

\b\b 
\ce{\bf 3.4. Heat loss and approach to equilibrium}

We 
believe that the loss of heat by radiation can be accounted
for by 
enlarging the system to include the electromagnetic field.
An 
approach of a closed system to equilibrium, to the extent that 
it
involves nothing more than a normalization of the temperature, 

may be encompassed by this, but it is clear that there are 
situations
where the electromagnetic field intervenes in more subtle 
ways, if at all.

For an example of special interest to this study 
consider an ideal gas in
complete isolation, in irrotational motion. 
Some of the energy may be
assotiated with this motion, and some with 
heat. Conceivably, after a
long time, the motion ceases and the 
kinetic energy becomes zero, while
the energy balance is assured by a 
rise in the temperature.

What we believe actally happens is this. In 
certain cases the motion does
not stop. If the vessel is a perfectly 
smooth torus with constant cross
section it is possible to imagine a 
perfectly uniform and
stationary circulating current.

\b\b

\b\b
\ce{\bf 3.5. Conclusions}

It seems that we have reached a 
conclusion, provisionally at least,
with respect to the influence of 
solar radiation on the atmosphere of the
earth. Though we are unable 
to describe the more or less uniform warming
of the atmosphere that 
certainly takess place, we are struck by the
difficulty of assigning 
the existence of a temperature gradient to the
incoming 
radiation.

This conclusion, if true, is shocking, for it contradicts 
a central
postulate of thermodynamics, according to which the parts 
any isolated
system, in eqiuibrium and in contact with each oter, 
must have the same
temperature. It is not thermodynamics in the 
narrowest sense that is being
questioned, but its application in the 
presence of gravity.

For the theory of atmospheres of moderate 
temperature this conclusion
would imply that the Lane-Ritter 
polytrope is an unexpectedly successful
inspiration.

Another 
surprising, tentative conclusion is that sources of
radiation are too 
weak, at least in the case of atmospheres similar to
that of the 
earth, to deflect the value of $k$ substantially away from
the 
critical value 3. It is therefore intresting to return to 
the
difficulty first raised in the Remark in Section 2.6. Namely, 
when the
action (2.10) is minimized with respect to both $\rho$ and 
$T$, we semm
to get too little freedom, too few equilibrium 
configurations.  Let us
examine this problem; thus we take the 
lagrangian (2.10), with $\mu$
as in (A.7), and in the notation 
(A.5),
$$
{\cal L}[\Phi,\rho,T] = \rho(\dot\Phi - \vec v^2/2 -gz + 
\lambda )
   -{\cal R}T\rho\log k   .~~ k = \rho/T^n 
\alpha_0.\eqno(2.10)
$$
The equations of motion are, the equation of 
continuity, and
$$
\dot\Phi - \vec v^2/2 -gz + \lambda = \R\rho( \log 
k +1),~~ \log k = n;
$$ the first from variation of $\rho$, the other 
from variation of $T$.
Note that this equation simply fixes what 
Ritter has called the
polytropic temperature.

\bb

\ve

\ce{\steptwo IV. Radiation}

The form (4.1) suggests the following action for a photon gas
($\epsilon = 1$)
$$
{\cal L}_{\rm ph} =   {\sigma\over 2}  F^2 -  W[\sigma],\eqno(4.3)
$$
where
$$
F^2 = {-1\over 2}g^{\mu\nu}g^{\alpha\beta}F_{\mu\alpha}F_{\nu\beta} =
\vec E^2 - \vec B^2.
$$
   This is not much better than a guess, but there is some justification for
it.

\ce{\bf 5.2. Justification for the photon gas lagrangian }

In favor of the choice (4.3) we advance the following.

      	a) It has the same structure as (4.1).

				b) It is covariant and gauge invariant.

	   c) We have concluded  that the photon gas must be
accompanied by an electromagnetic field. This is surprising but easy to
understand. In the case of an atomic gas there is a variable that is
canonically conjugate to the density; it is the velocity or in our case
the velocity potential; it is needed to formulate dynamics. The
electromagnetic field plays the same role for the photon gas, it is a
direct analogue of the velocity of the atomic gas. In the case of
equilibrium it is constant,  at least $F^2$ is constant. The direction
is fixed by spontaneous symmetry breaking, in practice it is probably a
random variable unless the gas is polarized by an external agent. If
correct, this feature of the theory may be most significant. It predicts
a Seebeck effect (Seebeck 1828) (Saha 1935) (page 29, thermocouples), not
only in metals and in dielectrics, but in vacuum as well.

     Next, let us take $W = b\sigma^2$.   In that case there is more.

~~d) Variation with respect to the density field $\sigma$ gives
$2b \sigma = F^2 $; substitution of this into the action
converts the potential term to a non-linear term of the Born-Infeld type.
Such terms are often invoked, for example, to calculate the Casimir
effect (Roberts 1983), to explain non-linear effects observed with
lasers (McKemma and Platzman 1963),
and to promote relaxation in the photon gas. Quantum electrodynamics also
predicts the presence of such terms in the effective lagrangian, though it
predicts a correction of the type
$(F\tilde F)^2$ as well, an indication that the lagrangian (4.3) may need
refinement. According to Euler (1935) and Karplus and Neuman (1950),
the effective action is
$$
-F^2 + c_1 (F^2)^2 + c_2 (F\tilde F)^2\eqno(4.4)
$$
with $c_1 = 
(5/80)\alpha^2/m^4, c_2 = -(7/90)\alpha^2/m^4$, where $\alpha$
is the 
fine structure constant and $m$ is the mass of the electron.
More 
about this below.
\bb

~~e) Let us consider a situation where the 
magnetic field is
effectively zero. The term $\vec H^2$ in the 
hamiltonian density  $\vec
E^2 + \vec H^2$ is    sometimes referred 
to as the kinetic part of the
energy, so that there may be some 
justification for viewing the case that
the magnetic field vanishes 
as an analogue of the static case in ordinary
gas dynamics. In that 
case the energy density is
$$
T_t^t =  -2\sigma F^2 
-p,\eqno(4.5)
$$
while the pressure, on shell, is
$$
p =-\sigma F^2 - 
W.\eqno(4.6)
$$
The equation of motion makes $-\sigma F^2 = 2  W$, so 
that $p = W$ and
$T_t^t = 3W$,
giving the correct equation of state 
for a photon gas, namely
$$
T_t^t = 3p.
$$
It is granted that getting 
this result in the case that the magnetic
field vanishes, and only in 
that case, adversely affects the strength of
this argument. On the 
other hand, any general relation between the scalar
field $p$ and the 
energy density would violate Lorentz invariance.

f) Since the energy 
density is proportional to the fourth power of the
temperature 
(Stefan-Boltzmann law) we must have $\sigma \propto T^2$.
The photon 
density $n$ is proportional to $\sigma^{3/2}$ and thus $p
\propto 
n^{4/3}$. All this is satisfactory.

g) The phenomenon of sound 
propagation in the photon gas deserves a
section of its 
own.

\b

\ce{\bf  4.3. The propagation of sound}

It is believed 
that the photon gas is capable of transmitting sound, at a
speed of 
$c/ \sqrt 3$  or less. An explanation for this is the
analogy with 
the dynamics of sound propagation in a polytropic gas. A
static 
solution of Eq.s (1.1-2) is $\rho = 1,  \vec v = 0$. To
   first 
order in $ \rho-1$ and $\vec v$ a sound wave travelling in the
$z$ 
direction is described by
$$
    \dot\rho + v' = 0,~~ -\dot v  = p' = 
\gamma p/\rho.
$$
hence $  \ddot \rho =  \gamma (p/\rho)  \rho''$ 
(Saha 1935, page 95).
The square of the speed
of propagation $\gamma 
p/\rho = dp/d\rho$.  Here
$\rho$ is essentially the energy density 
and thus  by analogy,
$dp/d\rho= 1/3$ for the photon gas.

It is 
important to be aware of the fact that this argument is based
on 
thermodynamics assisted by hydrodynamics. More precisely, it is
based 
on an analogy with an ordinary gas for which dynamical
hydrodynamics 
is an established theory. In the case of photons we readily
accept 
that the thermodynamic aspects are understood but, as far as we
know, 
an independent theory of photon hydrodynamics has not been
developed. 
However, the result is established by a standard
thermodynamical 
argument (reference).

Let us consider the equation of motion 
associated with (4.3).
Normalization constants do not interfere, so 
we continue to suppress the
factor $1/16\pi$. The equations are 
then
$$
2b\sigma = F^2, ~~ \p^\mu  \sigma (A_{\nu,\mu} - 
A_{\mu,\nu})=0.
$$
   In the gauge $A_0 = 0, ~~ $div$ \vec A = 0$ the 
first equation becomes
$$
2b\sigma = \sum_i \dot A_i^2 - 
\sum_{i,j}(\p_jA_i)(\p_jA_i).
$$
We consider the ``static" solution 
in which $  E_i = \dot A_i$ is
constant and of unit length,  $\sigma 
= 1/2b$, and
first order deviations $\d \sigma, \d A_i$ from it. In 
that case the first
equation gives
$$
b \d \sigma = \sum_j\dot A_j 
\d \dot A_j.
$$
   The second equation reduces to $\p^\mu \sigma 
\p_\mu A_i = 0$ or
$$
\d \dot \sigma \dot A_i + \sigma ( {\p^2\over 
\p t^2}  -
\Delta)\d  A_i = 0,
$$
whence
$$
   \d  \ddot {\sigma ~} 
\dot A_i + \sigma ( {\p^2\over
\p t^2}  -
\Delta)\d  \dot A_i = 
0,\eqno(4.7)
$$
Projected on $\dot A_i$ (and multiplied by 2) it 
becomes (since $2b\sigma
= 1$ and $\sum \dot A_i\dot A_i = 
1$)
$$
-2{\p^2\over \p t^2}\d  \ddot{\sigma~} = 2\sigma( {\p^2\over 
\p t^2}  -
\Delta)\sum_i \dot A_i\d \dot A_i=  2\sigma ({\p^2\over \p 
t^2}  -
\Delta )b\delta\sigma = ({\p^2\over \p t^2 }-\Delta)\d 
\sigma
$$
or $(3\p_t^2 - \Delta)\d \sigma = 0$. So finally the speed 
of the wave is
$ 1/\sqrt 3$. The direction of propagation is 
perpendicular to $\vec E$.

This happy result gives us some 
confidence in the lagrangian (A.3). With
numerical factors restored 
it is
$$
{\cal L}_{\rm ph} =  {\sigma\over 8\pi}  F^2 - b\sigma^2,~~ 
\eqno(4.8)
$$
The value of the constant $b$ unknown, so far, but 
since the energy
density is  $3b\sigma^2$ we know from the 
Stefan-Boltzmann law that
$$
3b\sigma^2 = \kappa T^4,~~ \kappa = 
7.5607 10^{-15} {\rm erg/cm^3
deg K}.\eqno(4.9)
$$
\b
\ce{\bf  4.4. 
Refinement of the lagrangian}

We return to Euler's result (A.4) for 
the non linear lagrangian but
ignore the last term for the time 
being.

As is standard practice, the symbol
$$
F^2 =\vec  E^2 -\vec 
B^2
$$
stands for an energy density or for the energy in a volume 
of
$1cm^3$; everything expressed in cgs units. Euler's formula (1936) 
is
$$
{\cal L} = F^2\Big(1+ {\alpha^2\over 
45\pi}({\hbar\over
m_ec})^4{F^2\over \hbar c}\Big).\eqno(4.10)
$$
All 
the factors are dimensionless, $\alpha$ is the fine 
structure
constant. Wiith $F^2$ it is
$$
{\cal L} = F^2\Big( 1 + 
{1\over 2.653326\times 10^6}({1\over
2.5913\times 
10^{10}})^4
{F^2\over 3.1638\times 10^{-17}}\Big) = F^2\Big( 1+ 
{F^2\over 3.785
\times 10^{31}}\Big).
$$
The correction is of the 
order of unity when $F^2$ gets up to
$10^{31} erg$ or about 
$10^{17}kWh$, a field strength of about
$6\times {15} gauss$, as seen 
only in pulsars. The value of $F^2$ at
the classical edge of the 
electron is about $10^{14} erg$.

For this to agree with our 
lagrangian  we have to include the
standard Maxwell lagrangian; thus 
we propose to modify (2.6), setting
instead
$$
{\cal L}_{\rm ph} = 
{1+\sigma\over 8\pi}  F^2 - b\sigma^2,~~ \eqno(4.11)
$$
The need to 
do so is clear once we identify $F$ with the
electromagnetic field 
strength. But we must review the evidence that
was presented in favor 
of (A.3), to see if it can be reconciled with
(4.11).

1. Perhaps 
one may be justified in speaking of 2 contributions to the
potential. 
The random fluctuations that are usually invoked to
explain the 
photon gas are normally summarized in the density
$\sigma$.
The usual 
treatment does not invoke a potential, and neither have we
done so up 
to now. But there are situations in which a global and
coherent field 
enters upon the scene as well, and then the ordinary
Maxwell action 
has to be included. It requires only a little positive
thinking to 
combine both in a single
electromagnetic potential. Thus one should 
ignore the term 1 in
$1+\sigma$ when dealing with the photon gas in 
the absence of a
global field.

2. If we were to replace $\sigma$ by 
$1+\sigma$ in the discussion of
sound propagation, in Eq.(4.7),
one 
would include that the speed is not $1/•\sqrt 3$ but that this as
the 
lower limit. That suggests that
this value would be found to apply in 
special situations only. We do
not know what the experimental 
situation is, except that the speed is
much less if a normal gas is 
present. Perhaps it will turn out that
attempts to excite a sound 
signal will result in light signals
instead.

3. Attention was called 
to the fact that the on shell value of the
lagrangian density 
coincides with the pressure, in hydrodynamics, and
also in Tolman's 
phenomenological treatment of relativistic
thermodynamics. But this 
leads to strange results in the context of
electromagnetism. The 
lagrangian,
in the case of a coherent beam of freely propagating 
photons,
vanishes. So, therefore, does the  of the pressure.
But the 
Pointing vector does not vanishes, and in the case of the
photon gas 
``the pressure" is found
by placing an imaginary wall in the way of 
the photons and assuming
that they are reflected or absorbed by the 
wall. It would be more
accurate to speak of  ``the pressure on the 
wall"; clearly there are
two quite different concepts of pressure. It 
is the pressure on the
wall that is used to derive the ratio of 3 
between energy density and
pressure, not "the pressure of the gas" in 
the sense of hydrodynamics.

In fact, a free photon by definition is 
not interacting with
anything. Free photons moving in a cavity, with 
or without an atomic
gas in it (besides the photon gas) can have no 
effect on the gas, or
they would not be free; in fact the quantity 
$F^2$ is a measure of
how much the photons are not free; that is, to 
what extent they
remember that they have been, or foresee that they 
shall be, in
interactions with the gas.

In conclusion, we  concede 
that  some of the arguments advanced in
favor of the original 
lagrangian (4.) lose some of their force as we
switch to (4. ). But 
the comparison with Born-Infeld theory compels
it.

\b

\ce{\bf 
4.5. Reflections on the program}

This theory is intended to provide 
a theory of heat, through the
identification of the ``photon energy 
density" $3b\sigma^2$ with $aT^4$,
the Stefan-Boltzmann law. On the 
way to achieving that goal it is
proposed to develop a dynamical 
theory of the photon gas.

The dynamical variable of the photon gas 
certainly must include some
density, energy density or photon number 
or some other density
related to one or both of these. We have seen 
that there are grounds for
identifying $b\sigma^2$ with the energy 
density in the sense
of thermodynamics. This identification may be 
only half right, but that is
not the problem that we wish to discuss 
here.. The question is whether
some additional field is needed. More 
precisely: is the appearance of an
electromagnetic field in this 
model an asset or is it an embarrasment?

The traditional treatment 
of ordinary gasses has served as a paradigm.
There too the density is 
an important dynamical variable, but the
formulation of a dynamical 
theory, besides the physical requirements,
are strong reasons to 
include the velocity field as an
additional, independent, dynamical 
variable. The traditional theory of
heat deals, in the first 
instance,  with the density of heat, a scalar
field, or equivalently 
with the temperature field, another scalar field.
In fact, the 
temperature is, sometimes, treated as a dynamical variable.
See e.g. 
Stayukovich 19
There is also a concept of heat flow, a vector field, 
but it is not an
independent dynamical variable, being related to the 
gradient of the
temperature. In laminar gas dynamics the velocity is 
the gradient of a
scalar field, but this scalar field is independent 
of the density field.

  From the point of view of the traditional 
theory of heat there is nothing
to suggest that an additional 
dynamical variable is needed. But in the
context in which we find 
ourselves the situation is quite different.
In the first place, we 
wish to give the theory a relativistic formulation,
and an action 
principle. The only relativistically invariant wave equation
for a 
scalar field is the Klein-Gordon equation, with some 
non-linear
generalizations. It is not impossible that a relativistic 
theory of heat
can be developed along these lines, but it would not 
be completely
satisfactory since it would have no apparent connection 
to
electromagnetism, the carrier of all heat flow.  In the second 
place we
need a coupling of the dynamical variables of heat to matter 
variables.
We know that the true, microscopic phenomenon responsible 
for heat
transfer is electromagnetic radiation, and we also have very 
definite
ideas as to the way that this radiation interacts with 
matter. It would seem
that this information must be taken into 
account, and that requires that
the electromagnetic field appear as 
one of the dynamical variables.
Finally, just as the velocity of flow 
needs to be defined to fully
describe the state of a gas, it may be 
thought that something besides
density may be required to describe 
the state of a photon gas. In
principle, a collection of photons is 
described by polarization and
wave number, both are lost in the 
amorphous concept of ``quantity of heat".

The analogy with ordinary 
gasses suggests an additional variable,
besides the density, and the 
origin of heat strongly suggests that this
variable be the 
electromagnetic potential.

But this brings us immediately to a 
difficulty, what  is
the direction of propagation and  polarization 
of
this radiation? Let us first suppose that the gas is polarized, 
all the
photons traveling in the same direction, a situation that may 
be
approximated in the case of radiation from a distant source.  In 
this case
it is natural to think that the electric field is oriented 
in the same
sense, and that a potential difference can be detected 
between two points
of the beam. An anologous effect is in fact 
observed in the case of heat
conduction in metals, it is the Seebeck 
effect (Seebeck 1821).

We come now to a characteristic feature of 
our model, and of similar
models, the relation $2b\sigma = -F^2$. If 
$F$ is the field of a set of
collimated, polarized photons, then $F^2 
= 0$. Such a beam does not by
itself imply temperature. Heat would 
appear, not because of the presence
of free photons, but because of 
scattering and absorption; that is, the
fact that in reality the 
photons are not free. Such scattering can take
place in a normal gas 
and it is expected to take place in a photon gas as
well, but so far 
we do not have a photon gas in which to verify this.

To generate 
temperature there are two mechanisms: we may prepare a beam
of off 
shell photons, which implies the existence of a source, or we 
may
introduce a degree of incoherent dispersion in the polarization 
or the
direction of propagation. Let us consider the second 
possiblity first.
Dispersion will make $F^2 \neq 0$ and the relation 
$2b\sigma =
-F^2$ may imply a variable temperature. But this 
relation
depends on the degree of dispersion, not on the  amplitude 
of the
field, so we are not able, at this point, to see any relation 
between the
strength of the field and the associated temperature. On 
the
other hand, once $\sigma$ is not constant the photons no 
longer
satisfy Maxwell's equations, they are not free, so that 
brings
us to the first possiblity. There is boot strap operating:
If 
$\sigma$ is not constant then the solutions for the potential
do not 
satisfy $F^2 = 0$.

Finally, in order to understand the nature of 
the field we have to solve
the equations of motion. Only then shall 
we be able to decide if it is
measurable and relevant to the theory 
of heat.

\b

   \ce{\bf   5.6. Radiation and heat in a hot 
gas}

Our aim is to calculate the heat generated in the photon gas by 
a hot body.
\b
1. Consider the static field generated by a charged 
particle at the origin
   of the coordinate system.
The potential 
will be assumed to have the form
$$
\vec A = 0,~ A_0 = 
f(r).
$$
Thus
$$
\vec E = {\rm grad}\, f = {\vec  x \over r}f' ,~~ 
\vec H = 0.
$$
and
$$
\sigma = -F^2 = \vec E^2 =  f'\,^2.
$$
The 
equation that we wish to solve is
${\rm div }(1+\sigma){\rm grad }f = 
0$, or
$$
r^2(1+f'\,^2)f' = {\rm constant}.
$$
We shall make use of 
the function
$$
\chi: x \mapsto (1+x^2)x
$$
and the  unique, real 
valued,  inverse function $\chi^{-1}$. Thus
$$
f'(r) = 
\chi^{-1}({c\over r^2}) ,~~ c = {\rm constant}.
$$
This function 
falls of as $1/r^2 $ at infinity and behaves
as $1/r^{2/3}$ near the 
origin.

The presence of the photon gas softens the singularity.
It 
acts as an infrared regulator,
just like the soft photons of quantum 
electrodynamics. It is as if the
   charge were spread out and less 
concentrated at a point.

The temperature is proportional to $E_z$ 
and thus it falls off as $1/r^2$
   at ``large" distances. Of course, 
in the case of an elementary charge we
expect that this temperature 
is extremely low and undetectable. There is
a non-zero energy density 
but no flow. Since there is no magnetic field
the energy density is 
exactly three times the pressure. This pressure has
nothing to do 
with the Pointing vector that  in this case vanishes.

\b

2. Next, 
let us consider a fixed dipole, consisting of a positive charge
at 
the origin and an opposite charge at the point $\vec d$ on 
the
$z$-axis.   The field equation is  a linear relation 
between
$(1+\sigma)F$ (not $F$) and the current, and the correct way 
to combine
the effect of the two sources is
$$
(1+\sigma)F 
=(1+\sigma_+)F_+ + (1+\sigma_-)F_-.\eqno(2.11)
$$
To first order in 
$d$ it is, schematically at first,
$$
(1+\sigma)F = d {\p\over \p 
z}(1+\sigma_+)F_+.
$$
Here $F_+$ has only one non-zero component,
 
$(F_+)_{zt} = (E_+)_z =: E_{+z}$, so the total field has the 
same
property, hence the precise statement is that
$$
(1+\sigma)E_z = 
d {\p\over \p z}\big(1+{E_{+z}}^2\big)E_{+z}=
-2dc {z\over 
r^4}.
$$
Notice that we are not making direct use of the result 
obtained previously
for the fields $\vec E_\pm$ of the individual 
charges. It is the sources
that are being added, as expressed by 
(2.11). Also, since $\sigma = -F^2$,
the last equation determines 
both $F$ and $\sigma$. Here $\sigma = E_z^2$,
so
$$
E_z = 
\chi^{-1}(-2dc {z\over r^4}).
$$
Notice that angular distribution of 
the field of a dipole is distorted
   along with the dependence on 
distance. The radiation field is therefore
not exactly a pure dipole. 
Of course, beyond the immediate neighborhood of
the dipole the effect 
is extremely small.
\b

3. An atomic dipole oscillates. The emitted 
radiation is predicted by
   atomic physics. It consists of photons, 
uncorrelated bursts of
monochromatic radiation with random phases. 
The field emitted by a
dipole
located at the origin has an electric 
component as well as
a
magnetic component.

Let us suppose that the
 
radiation is due to an
atomic, harmonic oscillator with fixed 
frequency
over a period of
several oscillation periods. Then it is 
the source and
not the field
that is monochromatic. The response is 
the field $(1+\sigma)
F$ and
it is 
satisfies
$$
\p^\mu(1+\sigma)F_{\mu\nu} \propto  \sin\omega
t,~~ \vec 
x\neq 0.
$$
Consequently,
$$
(1+\sigma)F_{\mu\nu}\propto
\sin\omega 
t,
$$
$$
(1+\sigma)^2\sigma^2 \propto\sin^2\omega
t,
$$
and
$$
\sigma 
= \chi^{-1}( c\sin\omega t),~~ c~{\rm independet~
of ~ time}.
$$
Thus 
a small amount of line broadening takes
place.

There is no radiating 
monopole, so we cannot construct the
oscillating
  dipole from a pair 
of oscillating charges. This makes
the calculation
more difficult and 
we shall be content to calculate
$\sigma $ to  the first
order of 
approximation. In that case we can
ignore the contribution 
of
$\sigma$ in the calculation of the fields,
to obtain the fields, 
and
finally $\sigma$, up to corrections of
order $\sigma^2$.

  There 
is no gauge in which $\vec A = 0$. There is
a gauge in which
  $A_0 = 
0$, but the simplest construction is as
follows. Suppose $\vec A$
is 
paralell to the third axis, and use the
Lorentz gauge, so that (up to 
a
factor that represents  the dipole
moment):
$$
A_3 = 
{\sin\omega(t-r)\over  r}, ~~{\rm div} \vec A
= -
{\omega z\over r^2} 
\cos (t-r)- {z\over
r^3}\sin\omega(t-r),
$$
$$
A_0 =  {  z\over 
r^2}\sin\omega
(t-r)-
{z\over \omega 
r^3}\cos\omega(t-r)
$$
and
$$
E_1 = -
{\omega\over r}{zx\over 
r^2}\cos\omega(t-r)
- {3\over 
r^2}{zx\over
r^2}\sin\omega(t-r)
+{3\over\omega 
r^3}{zx\over
r^2}\cos\omega(t-r),
$$
$$
   E_3 =  {\omega\over 
r}\sin^2\theta
\cos\omega  (t-r)
+  {1\over 
r^2}(1-3\cos^2\theta)\sin\omega(t-r)
  +{1\over \omega 
r^3}(3\cos^2\theta -1) \cos\omega(t-r),
   $$
   $$
  {\vec E}^2 = 
{\omega^2\over r^2}
\sin^2\theta\cos^2\omega(t-r)
+{2\o\over 
r^3}
\sin^2\theta\cos\o(t-r)
   \sin\o (t-r)
   $$
   $$
+ 
{1\over
r^4}(1+3\cos^2\theta) \sin^2\o (t-r) 
+{2\over
r^4}
\sin^2\theta\cos^2\o (t-r).
$$
Further, the magnetic 
field
is
$$
H_1 = -{\omega y\over r^2}  \cos\omega(t-r)
-{y \over 
r^3}
\sin\omega(t-r),
$$
$$
\vec H^2 = {\omega^2\over 
r^2}
\sin^2\theta\cos^2\omega(t-r)
+  {2\o\over 
r^3}
\sin^2\theta\cos\o(t-r)
   \sin\o (t-r)  + {1\over 
r^4}\sin^2\theta
\sin^2\o(t-r).
$$
There is a remarkable cancellation 
in
$$
   \vec E^2
- \vec H^2 = {2\over r^4}\cos^2\theta
+ {2\over 
r^4}[\sin^2\o(t-r) -
\cos^2\o(t-r)].
$$
This implies that the field 
rapidly approaches a
free field as the distance from the source 
increases.
Averaging over
several time periods and restoring a factor 
$d^2$ that
represents the
dipole moment density,  we
get to leading 
order (Hertz 1889)
$$
E^2 +
H^2 = d^2{ \omega^2\over 
r^2}\sin^2\theta,~~  \vec E^2 - \vec H^2
=
d^2{2\over 
r^4}\cos^2\theta, ~~ \vec E\wedge\vec H =
-{\omega^2\over r^3}\vec 
r\sin^2\theta\cos^2\omega(t-r).\eqno(2.6)
  $$

We try to estimate 
both quantities in a large volume of
homogeneous gas. The 
energy
density gives a divergent integral, but
we cut it off at
the 
mean free path $r_+$ to get the energy density
(at the origin 
and
thus everywhere)
$$
{\cal E} = d^2\omega^2\int
drd\Omega 
\sin^2\theta = {8\pi\over
3}d^2\omega^2\int_0^{r_+}dr
={4\pi\over 
3}d^2\omega^2r_+.
$$
We have cut off the divergent
integral at $r = 
r+$, where $r_+$ is the mean
free path of photons of
the given 
frequency in the gas.

For the integrated $\sigma$ we 
have
instead
$$
F^2 = 16b\sigma = 2d^2\int{drd\cos d\Omega 
\over
r^4}\cos^2\theta =
{8\pi\over 3}d^2\int {dr\over r^2} = 
{8\pi\over
3}d^2/r_-.
$$
Here we have an integral that diverges at 
the origin,
but we have seen
that the exact solution is less 
singular, so we have
cut the integral at
the effctive screening 
distance $r_-$.

Let us
assume, for simplicity only,  that our gas 
behaves like a black body,
then the
Stefan-Boltzman law tells us 
that
$$
{\cal E} = {4\pi\over
3}d^2\omega^2r_+ =  a 
T^4.\eqno(2.7)
$$
The Wien displacement
law, valid for normal 
temperatures, says that the
peak of the
frequency spectrum is 
at
$$
\omega =T/ \nu,~~\nu ~{\rm
constant}
$$
and this leads 
to
$$
{4\pi\over 3}{d^2\over \nu^2}r_+ =
  a T^2.
$$
We can use this 
result to express $\sigma$
as
$$
16b\sigma = {2\kappa \nu^2 \over 
r_-r_+} T^2.
$$
So
finally,
$$
\sigma = \xi T^2.\eqno(2.8)
$$

Values 
of the various
factor are
$$
\kappa =  7.56410^{-15} 
erg/cm^3\,^o\hskip-.9mm K^4
\nu = .28977/cm
$$
$$
\xi  = 1.27 \times 
10^{-15}r_+r_-
\eqno(2.9)
$$
Values of the distances $r_{\pm}$ are 
lacking, though
it is not too hard to
give the order of magnitude of 
$r_+$. The
theory suggests that the product
is nearly the same for 
all gasses
with a high coefficient of absorption.

In the same manner 
we shall
calculate the Pointing vector. Averaging over time we 
get
$$
\vec
E\wedge \vec H = {-1\over 2} \sin^2\theta~{\omega^2\over 
r^3}\vec r.
$$
This is the field at $x$ due to a dipole at the origin, aligned with
the $z$-axis. So we have a radial field
of strength ${-1\over 2} \sin^2\theta~{\omega^2\over  r^2}$.
Averaging over the orientation of the dipole
yields $(-1/3)(\omega^2/r^2)$. This is also the field at the origin
due to a dipole at $x$, hence
$$
\vec E\wedge \vec H|_0 = {1\over 3}\int{\omega^2\over r^3}\vec r d^2(x)d^3x.
$$
The main distribution comes from a regioin where we can approximate
$$
d^2(x) = d^2(0) + \vec r\cdot{\rm grad}~ d^2(0),
$$
The integral, cut off at $r = r_+$  gives
$$
\vec E\wedge \vec H|_0 = {2\pi\omega^2\over 9}r_+^2{\rm grad}d^2(0)
$$
See notes to finish.
\ve

\b

\b

\b\b
\no{\bf References}

\no Bernoulli, D., ~ Argentorat,
1738.

\no Chandrasekhar, S.

\no Euler, H.,   -ber die Streuung von
Licht an Licht nach der Diracschen Theorie,

\quad Ann.Phys.  {\bf
26} 398-  (1936).

\no Fronsdal,  C.,  Ideal Stars and General
Relativity, to appear in Gravitation and

\quad General Relativity,
gr-qc/0606027.

\no  Fronsdal, C., Stability of polytropes,
arXiv
0705.0774 [gr-cc]

\no  Davidson, R.D., ``Theory of Non-Neutral
Plasmas",
Addison-Wesley
    1990.

\no Karplus, R. and Neuman, M.,
Non-linear Interactions between Electromagnetic Fields,

\quad
Phys.Rev. {\bf 80} 380-385 (1950).

\no Mazur, P. and Mottola, 
E.,
Gravitational Vacuum Condensate
Stars,

\no 
\quad
\quad~~~gr-qc/0407075

\no McKenna, J. and Platzman,  P.M., 
Nonlinear
Interaction of Light in Vacuum,

\quad Phys. Rev. {\bf 
129}
2354-2360 (1962).

\no Milne

\no Rees, M.F., Effects of very 
long
wavelength primordial
gravitational 
radiation,

\quad
Mon.Not.astr.Soc. {\bf 154} 187-195 (1971).

\no 
Saha,  M.N. and
Srivastava, B.N., {\it A treatise on heat}, The 
Indian Press, 1935.

\no Stanyukovich, K.P., {\it Unsteady motion of 
continuous media},

\quad Pergamon
Press New York 
1960.

\ve

\ce{\stepthree
Studies in 
Thermodynamics}

\no{\steptwo  I. Introduction}

1.1. 
Simple
hydrodynamics

1.2. Laminar flow

1.3. Variational 
formulation

1.4. On shell
relations and the potential

1.5. 
Equations of state
1.6.
The
mass

\no{\steptwo II The first 
law}

1.1. Thermodynamic equilibrium

1.2. The ideal gas in 
thermodynamics

1.3. The
ideal gas in statistical mechanics

1.4. The 
first law and the internal energy

1.5. The first law and the 
hamiltonian
\b
\no {\steptwo IV Radiation}

4.1. Lorentz 
invariance

Justification for the photon gas 
lagrangian

The
propagation of sound

Reflections on the 
program

Radiation and heat
in a photon gas
\b
\no {\steptwo III. A 
theory of temperature and
radiation}

\end{document}